\documentclass[
reprint,
superscriptaddress,
 amsmath,amssymb,
floatfix,
]{revtex4-2}

\UseRawInputEncoding
\usepackage[normalem]{ulem}
\usepackage{graphicx}
\usepackage{stackengine}
\usepackage{color}
\usepackage{dcolumn}
\usepackage{bm}
\usepackage{siunitx}
\usepackage{enumitem}

\makeatletter
\def\@email#1#2{%
 \endgroup
 \patchcmd{\titleblock@produce}
  {\frontmatter@RRAPformat}
  {\frontmatter@RRAPformat{\produce@RRAP{*#1\href{mailto:#2}{#2}}}\frontmatter@RRAPformat}
  {}{}
}%
\makeatother

\begin{document}



\title{Mg-doping and free-hole properties of hot-wall MOCVD GaN}

\author{A.~Papamichail}
\email{alexis.papamichail@liu.se}
\affiliation{Center for III-Nitride Technology, C3NiT - Janz\'en, Link\"{o}ping University, Department of Physics, Chemistry and Biology (IFM), SE-58183 Link\"{o}ping, Sweden}
\affiliation{Terahertz Materials Analysis Center, THeMAC, Link\"{o}ping University, SE-58183 Link\"{o}ping, Sweden}

\author{A.~Kakanakova}
\affiliation{Center for III-Nitride Technology, C3NiT - Janz\'en, Link\"{o}ping University, Department of Physics, Chemistry and Biology (IFM), SE-58183 Link\"{o}ping, Sweden}

\author{E. \"O. ~Sveinbj\"ornsson}
\affiliation{Center for III-Nitride Technology, C3NiT - Janz\'en, Link\"{o}ping University, Department of Physics, Chemistry and Biology (IFM), SE-58183 Link\"{o}ping, Sweden}
\affiliation{Science Institute, University of Iceland, Dunhagi 5, IS-105 Reykjavik, Iceland}

\author{A.~R.~Persson}
\affiliation{Center for III-Nitride Technology, C3NiT - Janz\'en, Link\"{o}ping University, Department of Physics, Chemistry and Biology (IFM), SE-58183 Link\"{o}ping, Sweden}
\affiliation{Thin Film Physics Division, Department of Physics, Chemistry and Biology (IFM), Link\"{o}ping University, Link\"{o}ping SE-58183, Sweden}

\author{B.~Hult}
\affiliation{Department of Microtechnology and Nanoscience, Chalmers University of Technology, SE-41296, G\"{o}teborg, Sweden}

\author{N.~Rorsman}
\affiliation{Department of Microtechnology and Nanoscience, Chalmers University of Technology, SE-41296, G\"{o}teborg, Sweden}

\author{V.~Stanishev}
\affiliation{Terahertz Materials Analysis Center, THeMAC, Link\"{o}ping University, SE-58183 Link\"{o}ping, Sweden}

\author{S.~P.~Le}
\affiliation{Center for III-Nitride Technology, C3NiT - Janz\'en, Link\"{o}ping University, Department of Physics, Chemistry and Biology (IFM), SE-58183 Link\"{o}ping, Sweden}

\author{P.~O.~\AA.~Persson}
\affiliation{Thin Film Physics, Department of Physics, Chemistry and Biology (IFM), Link\"{o}ping University, Link\"{o}ping SE-58183, Sweden}

\author{M.~Nawaz}
\affiliation{Hitachi Energy, Forskargr€and 7, 721 78 Vester\'as, Sweden}

\author{J.~T.~Chen}
\affiliation{Center for III-Nitride Technology, C3NiT - Janz\'en, Link\"{o}ping University, Department of Physics, Chemistry and Biology (IFM), SE-58183 Link\"{o}ping, Sweden}
\affiliation{SweGaN AB, Olaus Magnus v\"ag 48A
SE-58330 Link\"oping, Sweden}

\author{P.~P.~Paskov}
\affiliation{Center for III-Nitride Technology, C3NiT - Janz\'en, Link\"{o}ping University, Department of Physics, Chemistry and Biology (IFM), SE-58183 Link\"{o}ping, Sweden}

\author{V.~Darakchieva}
\email{vanya.darakchieva@liu.se; vanya.darakchieva@ftf.lth.se}
\affiliation{Center for III-Nitride Technology, C3NiT - Janz\'en, Link\"{o}ping University, Department of Physics, Chemistry and Biology (IFM), SE-58183 Link\"{o}ping, Sweden}
\affiliation{Terahertz Materials Analysis Center, THeMAC, Link\"{o}ping University, SE-58183 Link\"{o}ping, Sweden}
\affiliation{Solid State Physics and NanoLund, Lund University, P. O. Box 118, 221 00 Lund, Sweden}

\date{\today}

\begin{abstract}

The hot-wall metal-organic chemical vapor deposition (MOCVD), previously shown to enable superior III-nitride material quality and high performance devices,  has been explored for Mg doping of GaN. We have investigated the Mg incorporation in a wide doping range ($2.45\times{10}^{18}$ \si{cm^{-3}} up to $1.10\times{10}^{20}$ \si{cm^{-3}})  and demonstrate GaN:Mg with low background impurity concentrations under optimized growth conditions. Dopant and impurity levels are discussed in view of Ga supersaturation which provides a unified concept to explain the complexity of growth conditions impact on Mg acceptor incorporation and compensation. The results are analysed in relation to the extended defects, revealed by scanning transmission electron microscopy (STEM), X-ray diffraction (XRD), and surface morphology, and in correlation with the electrical properties obtained by Hall effect and capacitance-voltage (C-V) measurements. This allows to establish a comprehensive picture of GaN:Mg growth by hot-wall MOCVD providing guidance for growth parameters optimization depending
on the targeted application. We show that substantially lower H concentration as compared to Mg acceptors can be achieved in GaN:Mg without any in-situ or post-growth annealing resulting in p-type conductivity in as-grown material.  State-of-the-art $p$-GaN layers with a low-resistivity and a high free-hole density (0.77 $\Omega$.cm and $8.4\times{10}^{17}$cm$^{-3}$, respectively) are obtained after post-growth annealing demonstrating the viability of hot-wall MOCVD for growth of power electronic device structures. 

\end{abstract}

\pacs{Valid PACS appear here}

\maketitle

\section{Introduction}


Mg doping of GaN is essential for obtaining epitaxial layers with $p$-type conductivity for GaN-based devices such as light-emitting diodes (LEDs), normally-off high electron mobility transistors (HEMTs), and  $p$-$n$ diodes. Metal-organic chemical vapor deposition (MOCVD) is the technique of choice to grow large-scale GaN-based device structures and  $p$-type doping in MOCVD GaN:Mg  with free hole concentrations in the range of $10^{16}$ - $10^{18}$~cm$^{-3}$ has successfully been demonstrated. \cite{Kozodoy_2000, Tetsuo2018WideConcentrations, Narita_2020}

As-grown MOCVD Mg-doped GaN layers have limited or no $p$-type conductivity due to the formation of Mg-H complexes.\cite{GotzAPL95} Originating from the pyrolysis of hydrogen carrier gas and from the ammonia precursor cracking, atomic H is abundant in the MOCVD processes. The incorporated hydrogen must dissociate from the Mg-H complexes and out-diffuse to render GaN $p$-type, which requires post-growth treatment of the as-grown material. \cite{amano1989p, Nakamura_1992} The activation is typically achieved via ex-situ rapid or regular thermal annealing under  N${_2}$ atmosphere. Due to the large Mg acceptor ionization energy, the free hole concentration at room temperature is much lower than the Mg doping concentration. For this reason, high Mg concentration should be incorporated in GaN to reduce the ionization energy. \cite{Kozodoy_2000, brochen2013dependence} However, above a certain level, typically in the low 10$^{19}$ cm$^{-3}$ range, further increase of Mg concentration leads to a reduction in free hole concentrations. \cite{Kozodoy_2000,castiglia2011role,Narita_2020} The latter is often associated with the formation of pyramidal inversion domains (PIDs). \cite{PID_APL99,Tetsuo2018WideConcentrations} It has been shown that Mg atoms segregate at the PID (0001) boundaries rendering Mg atoms electrically inactive. \cite{Narita_2020} Other works have suggested that the nitrogen vacancy, V$_\mathrm{N}$ and its complexes (e. g. Mg-V$_\mathrm{N}$) play a role in the compensation of Mg acceptors at high dopant concentrations. \cite{WalleAPL12,PhysRevB.93.165207} In addition, any donors, e. g. unintentional impurities such as Si and O or native defects such as Ga interstitials, will interfere negatively with the desired $p$-type conductivity. Notably, C is a major impurity  in MOCVD due to the metal-organic precursor molecules cracking. The roles of C as a trap and compensation defect in GaN:Mg have been extensively discussed. Density functional theory calculations showed that C occupying Ga site, is a shallow donor and it has the lowest formation energy for Fermi level positions close to the valence band maximum. \cite{lyons2014effects}  Experimentally, it was demonstrated that the C incorporation leads to an increased donor concentration and a reduction in hole mobility, which was attributed to a donor-like +1/0 state of C$_\mathrm{N}$. \cite{Narita_2020} Variation of growth conditions, such as V/III ratio, precursor and gas flows, is expected to affect impurity incorporation and defect formation energies, and hence their densities, might influence the degree of passivation/compensation of Mg acceptors. Therefore, optimization and tuning of the growth process is explicitly necessary for acquiring low impurity  and native defect levels in GaN:Mg.

Continuous efforts in further improving $p$-type conductivity and free hole properties, including both growth and post-growth acceptor activation, are undertaken to  further the progress in GaN-based device performance. \cite{Narita_2020, klump2020control} Despite the numerous investigations, no study exists on the capability of hot-wall MOCVD to deliver high-quality GaN:Mg layers with $p$-type conductivity. Because of its concept and the range of high deposition temperatures (up to 1400 - 1600 $^{\circ}$C) achievable, the hot-wall MOCVD has successfully complied to conditions required for deposition of epitaxial layers of Al(Ga)N of semiconductor quality \cite{Kakanakova-Georgieva2009Hot-wallAlN, Schoche2017InfraredEffect} and reflected in demonstrating exciton luminescence in photoluminescence measurements \cite{Feneberg2015ExcitonAnnealing} and Mg-doped Al$_{0.85}$Ga$_{0.15}$N layers with low resistivity at room temperature. \cite{Kakanakova-Georgieva2010Mg-dopedTemperature} Recently,  a room-temperature mobility above 2200 cm$^{2}/\mathrm{V.s}$ of two-dimensional electron gas (2DEG) AlGaN/GaN HEMT heterostructures, \cite{Chen2015Room-temperatureHeterostructure, Armakavicius2016PropertiesEffect} a GaN$-$SiC hybrid material for high-frequency and power electronics \cite{Chen2018AElectronics} and state-of-the-art N-polar AlN epitaxial layers\cite{Zhang_2020n} using hot-wall MOCVD have been demonstrated. The hot-wall MOCVD concept enables reduced temperature gradients in both vertical and horizontal directions. It further allows for independent control of the gas phase chemistry over the substrate and the growth species surface diffusion, which may be exploited to control defect formation and impurity incorporation in a wider growth window.

In this work, we report a systematic study of Mg-doped GaN epitaxial layers grown by hot-wall MOCVD. A detailed investigation of the effects of growth and doping conditions on the incorporation of Mg, H and C impurities is presented and discussed in terms of Ga supersaturation. The results are evaluated in relation to extended defects, revealed by transmission electron microscopy as well as X-ray diffraction and surface morphology, and correlated with the electrical and free-hole properties of the GaN:Mg layers obtained by Hall-effect and capacitance-voltage measurements. As a result, a comprehensive picture of hot-wall MOCVD of GaN:Mg is established, demonstrating the capabilities of this technique to deliver state-of-the-art $p$-type GaN.

\section{Experimental details}
 
A hot-wall MOCVD reactor in horizontal configuration was utilized for the epitaxial growth of all layers. Chemical-mechanical polished 4H-SiC substrates with on-axis $(0001)$ orientation were used after wet chemical cleaning treatment. Before growth, the cleaned substrates were annealed and etched with hydrogen at 1340~$^\circ$C in the MOCVD reactor. AlN nucleation layers (NL) with thickness of $\sim$50~nm were grown at \SI{1250}{\celsius} with a V/III ratio of 1258, followed by growth of 500~nm-thick Mg-doped GaN layers. The growth process was performed under a constant ammonia (NH${_3}$) flow ($2~\ell$/min), a mixture of N${_2}$ and H${_2}$ as carrier gases and a constant pressure of $100$ mbar. NH${_3}$, Trimethylaluminum (TMAl), Trimethylgallium (TMGa) and bis(cyclopentadienyl)magnesium (Cp${_2}$Mg) were used as the N, Al, Ga and Mg precursors respectively.  
The growth conditions of the GaN:Mg layers are summarized in Table~\ref{tab:sample growth conditions}. The unintentionally doped (UID) GaN reference sample and the Mg-doped GaN samples with Mg doping concentration ranging from $2.45\times{10}^{18}$ \si{cm^{-3}} up to $1.10\times{10}^{20}$ \si{cm^{-3}}, are denoted as M$_0$ and M$_1$-M$_9$, respectively. Several sets of Mg-doped GaN layers were grown by systematically varying one of the following growth parameters: ($i$) Cp${_2}$Mg/TMGa ratio -  samples  M$_1$ - M$_5$ grown under optimized conditions with low carrier gas flows, where the Cp${_2}$Mg/TMGa  is varied between 0.033\% and 0.5\% and samples - M$_7$ and M$_8$ grown with high carrier gas flows with Cp${_2}$Mg/TMG of 0.335\% and 0.5\%, respectively, ($ii$) V/III ratio - samples M$_4$, M$_6$ and M$_7$ grown with Cp${_2}$Mg/TMGa of 0.335\% and V/III ratios of 906, 1811 and 453, respectively, and ($iii$) growth temperature (T$_\mathrm{g}$) - samples M$_8$ and M$_9$ with T$_\mathrm{g}$\,=\,1120~$^\circ$C and T$_\mathrm{g}$\,=\,1040~$^\circ$C, respectively.
In addition, a sample, M$_{\mathrm{opt}}$ was grown on 1 $\mu$m-thick undoped GaN using the conditions of M$_3$. The Mg acceptors were activated using optimized ex-situ annealing process at \SI{900}{\celsius} in N$_2$ atmosphere  
in a rapid thermal processing equipment.

\begin{table*}
\caption{\label{tab:Sample growth conditions} A summary of the growth conditions for the GaN:Mg layers studied in this work: growth temperature (T$_g$), growth pressure, V/III ratio, growth rate, Cp$_2$Mg/TMGa ratio, and H$_2$ and N$_2$ carrier gases  flows.} 
\begin{ruledtabular}
\begin{tabular}{{l}{c}{c}{c}{c}{c}{c}{c}{c}{c}{c}{c}}
Sample & T$_g$  & pressure & V/III ratio & growth rate & Cp$_2$Mg/TMGa & H$_2$ & N$_2$ \\
& ($^{\circ}$C) & (mbar) & & (\si{\mu m}$/$h) & ($\%$) & ($\ell$/min) & ($\ell$/min)  \\
\hline  
M$_0$ & $1120$ & $100$ & $906$ & $0.70$ & $0$ & $19$ & $9$ \\
M$_1$ & $1120$ & $100$ & $906$ & $0.70$ & $0.033$ & $19$ & $9$ \\
M$_2$ & $1120$ & $100$ & $906$ & $0.60$ & $0.082$ & $19$ & $9$ \\
M$_3$ & $1120$ & $100$ & $906$ & $0.70$ & $0.167$ & $19$ & $9$ \\
M$_4$ & $1120$ & $100$ & $906$ & $0.70$ & $0.335$ & $19$ & $9$ \\
M$_5$ & $1120$ & $100$ & $906$ & $0.60$ & $0.500$ & $19$ & $9$ \\
M$_6$ & $1120$ & $100$ & $1811$ & $0.14$ & $0.335$ & $19$ & $9$ \\
M$_7$ & $1120$ & $100$ & $453$ & $1.30$ & $0.335$ & $25$ & $12$ \\
M$_8$ & $1120$ & $100$ & $453$ & $1.06$ & $0.500$ & $25$ & $12$ \\
M$_9$ & $1040$ & $100$ & $453$ & $1.20$ & $0.500$ & $25$ & $12$ \\
\hline
M$\mathrm{_{opt}}$ & $1120$ & $100$ & $906$ & $0.60$ & $0.167$ & $19$ & $9$ \\


\end{tabular}
\end{ruledtabular}
\label{tab:sample growth conditions}
\end{table*}

The Mg and the background impurity (H, C, Si, O) depth profiles in as-grown and annealed samples were measured by secondary ion mass spectroscopy (SIMS). The detection limit was $1\times{10}^{16}$ \si{cm^{-3}} for Mg, $1\times{10}^{17}$ \si{cm^{-3}} for H, (3-5)$\times{10}^{15}$ cm$^{-3}$ for Si and O, and $1\times{10}^{15}$ \si{cm^{-3}} for C. The average concentrations for all the elements were estimated using the SIMSview program from EAG Labs. \cite{EAGLaboratoriesHttps://www.eag.com}  
The surface morphology of the layers was studied by atomic force microscopy (AFM) using a Veeco Dimension 3100 scanning probe microscope in tapping mode. Both $5\times{5}\,\mu$m$^{2}$ and  $80\times{80}\,\mu$m$^{2}$ images  were acquired. Spectroscopic ellipsometry measurements were performed on a J. A. Woollam RC$2$-XI ellipsometer in the ultraviolet-visible spectral range ($0.7-5.9~\si{\eV}$) for the determination of the layer thicknesses. The crystalline quality of the layers was evaluated by high-resolution X-ray diffraction (HRXRD) using a PANalytical Empyrean diffractometer. A hybrid monochromator consisting of a parabolic x-ray mirror and a two-bounce Ge($220$) crystals resulting in CuK$\alpha_1$ radiation with wavelength $\lambda$ = \SI{1.5405974}{\angstrom} at the incident x-ray beam and a symmetric three-bounce Ge($220$) analyzer on the detector side were used.  HRXRD $2{\theta}-{\omega}$ scans of the symmetric 0006 and the asymmetric $10\overline{1}5$ Bragg peaks were used for the determination of the $c$ and $a$ lattice parameters, respectively. 
The screw- and edge-type dislocation densities, $N_\mathrm{S}$ and $N_\mathrm{E}$, were estimated using the tilt and twist angles, $\alpha_S$ and $\alpha_E$, respectively \cite{Srikant1997MosaicMismatch, Metzger1998DefectDiffractometry} and the lattice parameters of a bulk GaN\cite{Darakchieva_2007} 

\begin{equation} 
\label{eqn:dislocation density}
\begin{split}
N\mathrm{_{S}} & = \frac{\alpha\mathrm{_S^{2}}}{4.35b_{c}^{2}} \\
N\mathrm{_{E}} & = \frac{\alpha\mathrm{_E^{2}}}{4.35b_{a}^{2}}
\end{split}
\end{equation}
The tilt angle was determined by the Williamson-Hall plots of the symmetric $0002$, $0004$, and $0006$ diffraction peaks of GaN, \cite{Metzger1998DefectDiffractometry} and the magnitude of the Burger vector along the $c$-axis ($b_c$ = 0.5185 nm) was used for an estimation of the screw-type dislocation density. For the edge dislocation density estimation, a method proposed by Srikant $et~ al.$, \cite{Srikant1997MosaicMismatch} where the full-width at half-maximum (FWHM) of the rocking curves from the asymmetric $10\overline{1}1$, $10\overline{1}2$, $10\overline{1}3$, $10\overline{1}4$ $10\overline{1}5$, and $30\overline{3}2$ diffraction peaks and the Burger vector along the $a$-axis ($b_a$ = 0.3189 nm) are used.

The structural quality of selected samples was further assessed by transmission electron microscopy (TEM) using the double corrected Link\"{o}ping FEI Titan$^3$ 60-300 microscope, operated in scanning TEM (STEM) mode and 300 kV. Images were acquired using annular dark-field (ADF) and annular bright-field (ABF) detectors (collection angles $\sim21-200$ and $\sim4-43$ mrad respectively), revealing both atomic number and strain contrast. Electron energy loss spectroscopy (EELS) thickness measurements were performed by acquisition using a Gatan GIF Quantum detector with $10$ ms exposure time, $0.25$ eV/channel dispersion ($-50.0$ eV to $462.0$ eV), and convergence and collection angles of $22.0$ and $15.0$ mrad respectively. Calculations were done in the embedded function in Gatan Digital Micrograph software. Samples were mechanically cut and mounted in Ti grids, followed by mechanical thinning and Ar-ion milling (Gatan PIPS Model 691). This produced two projection directions: $[1\bar{1}00]$ and $[11\bar{2}0]$.

The free-hole concentration and mobility of the as-grown and annealed samples were determined by Hall-effect measurements performed at room temperature in Van der Pauw configuration using a Linseis HCS1 instrument. For this purpose, Ni/Au ($5$ \si{nm}/$250$ \si{nm}) ohmic contacts were deposited by thermal evaporation and annealed at \SI{450}{\celsius} in air. Note that the Hall-effect measurements of as-grown and annealed samples were performed on the same piece for a given growth condition. The effective dopant concentration $N\mathrm{_A}-N\mathrm{_D}$ in samples M$_1$ - M$_8$ was extracted by capacitance-voltage C-V measurements, using a Hg-probe setup with a 4284A LCR meter from Agilent. The C-V data were acquired in series measurement mode in the 1 - 10 kHz frequency range.

The concentrations of Mg dopants, H and C impurity, the root mean square surface roughness, the density of screw and edge type dislocations and the free-hole concentration in as-grown GaN and the free hole concentration, mobility and resistivity of the respective thermally activated GaN samples are summarized in Table II.

\begin{table*}
\caption{\label{tab:table2} {Concentrations of Mg dopants ([Mg]), H ([H]) and C ([C]) unintentional impurities, root mean square (RMS) surface roughness, screw ($N_\mathrm{S}$) and edge ($N_\mathrm{E}$) dislocation densities, hole concentration ($p_0$) in the as-grown GaN:Mg layers, and hole concentration ($p$), mobility ($\mu$), and resistivity ($\rho$) in the respective layers after annealing.}}


\begin{ruledtabular}
\begin{tabular}{{l}{c}{c}{c}{c}{c}{c}{c}{c}{c}{c}{c}}
 & \multicolumn{7}{c} {as-grown} & & \multicolumn{3}{c} {annealed} \\
 \cline{2-8}  \cline{10-12}\\
 Sample & [Mg] & [H] & [C] & RMS & $N_\mathrm{S}$ & $N_\mathrm{E}$ & $p_0$ & & $p$ &  $\mu$  & $\rho$ \\
& $(\si{\centi\meter}^{-3})$ & $(\si{\centi\meter}^{-3})$ & $(\si{\centi\meter}^{-3})$ & $(\si{\nano\meter})$ & $(\si{\centi\meter}^{-2})$ & $(\si{\centi\meter}^{-2})$ & $({10}^{16}\si{\centi\meter}^{-3})$ & & $({10}^{17}\si{\centi\meter}^{-3})$ & ($\si{\centi\meter}^{2}/\mathrm{V.s}$) & $(\Omega.\si{\centi\meter})$  \\
\hline  
M$_0$  & $0$ & $1.5\times{10}^{17}$ & $9.9\times{10}^{15}$ & $0.15$ & $1.7\times{10}^{7}$ & $5.0\times{10}^{8}$ & - & & - & - & - \\
M$_1$  & $2.5\times{10}^{18}$ & $2.4\times{10}^{18}$ & $9.8\times{10}^{15}$ & $0.22$ & $3.8\times{10}^{7}$ & $5.9\times{10}^{8}$ & N/A & & N/A & - & - \\
M$_2$ & $6.0\times{10}^{18}$  & $4.8\times{10}^{18}$ & $6.0\times{10}^{15}$ & $0.22$ & $4.0\times{10}^{7}$ & $6.2\times{10}^{8}$ & $8.4$ & & $5.8$ & $10$ & $1.08$ \\
M$_3$ & $1.6\times{10}^{19}$  & $1.5\times{10}^{19}$ & $1.5\times{10}^{16}$ & $0.19$ & $2.1\times{10}^{7}$ & $5.4\times{10}^{8}$ & $ 0.76$ & & $6.5$ & $9$ & $1.08$ \\
M$_4$  & $2.4\times{10}^{19}$  & $2.2\times{10}^{19}$ & $1.5\times{10}^{16}$ & $0.19$ & $3.7\times{10}^{7}$ & $4.8\times{10}^{8}$ & N/A & & $5.3$ & $9$ & 1.30 \\
M$_5$  & $6.1\times{10}^{19}$  & $9.6\times{10}^{18}$ & $8\times{10}^{15}$ & $0.26$ & $2.9\times{10}^{7}$ & $8.1\times{10}^{8}$ & $9.6$ & & $2.3$ & $9$ & $2.90$ \\
M$_6$  & $1.6\times{10}^{19}$  & $1.3\times{10}^{19}$ & $5.3\times{10}^{15}$ & $0.17$ & $3.0\times{10}^{7}$ & $7.4\times{10}^{8}$ & $3.1$ & & $8.9$ & $8$ & $0.88$\\
M$_7$  & $7.0\times{10}^{19}$  & $1.4\times{10}^{19}$ & $2.1\times{10}^{16}$ & $0.20$ & $2.4\times{10}^{7}$ & $9.0\times{10}^{8}$ & $0.61$ & & $0.9$ & $9$ & $7.78$ \\
M$_8$  & $6.7\times{10}^{19}$  & $1.4\times{10}^{19}$ & $2.3\times{10}^{16}$ & $0.45$ & $5.4\times{10}^{7}$ & $7.0\times{10}^{8}$ & $0.02$ & & $1.1$ & $6$ & $8.73$ \\
M$_9$  & $1.1\times{10}^{20}$  & $2.9\times{10}^{19}$ & $3.1\times{10}^{17}$ & $1.00$ & - & - & N/A & & N/A & - & - \\
\hline
M$_{\mathrm{opt}}$  & $1.6\times{10}^{19}$  & $1.5\times{10}^{19}$ & $1.5\times{10}^{16}$ & $0.18$ & $1.6\times{10}^{7}$ & $4.3\times{10}^{8}$ & $1.67$ & & $8.4$ & $10$ & $0.77$ 

\end{tabular}
\end{ruledtabular}
\label{tab:sample properties}
\end{table*}

\section{Results and discussion}
\subsection{Mg incorporation}

\begin{figure}
\includegraphics[keepaspectratio=true,width=\linewidth, clip, trim=0cm 0cm 0cm 0cm ]{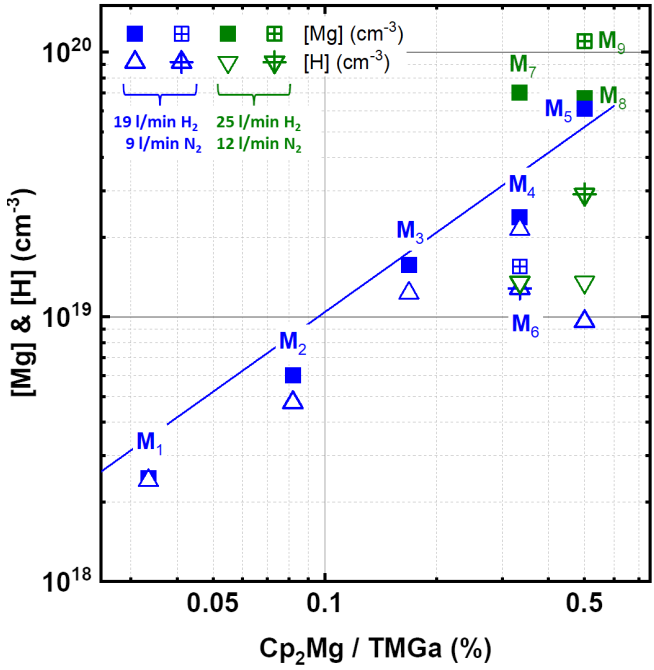}
\caption{Mg and H concentrations ([Mg] and [H]) determined by SIMS as a function of Cp$_2$Mg/TMGa ratio in the as-grown GaN:Mg layers. Squares and triangles represent [Mg] and [H], respectively. The sample labels according to Table I are indicated in the figure. The blue line represents a linear fit to the [Mg] data points for samples M$_1$ - M$_5$ grown under the same growth conditions.}
\label{fig:MgvsCpMg_TMG}
\end{figure}

Figure~\ref{fig:MgvsCpMg_TMG} shows the Mg and H concentrations in the GaN:Mg layers as a function of the ratio between the dopant precursor and the TMGa in the gas phase, Cp$_2$Mg/TMGa. As expected, Cp$_2$Mg/TMGa has a significant effect on the Mg incorporation. For samples M$_1$ - M$_5$ grown under optimized growth conditions of V/III ratio of 906  and growth rate of 0.60-0.70 $\mu\mathrm{m}$/h,  [Mg] increases linearly  with Cp$_2$Mg/TMGa. A maximum Mg concentration of ~${\sim{6.0\times{10}^{19}}}$ \si{{cm}^{-3}} at Cp$_2$Mg/TMGa = 0.5\% is reached.  We deduce a dopant incorporation efficiency of 0.24 from the dependence of the atomic fraction of magnesium [Mg]/[Ga] in the lattice as a function of the Cp$_2$/TMGa flux ratio in the vapour phase (see Fig. S1 in the supplementary material). The observed trends (Fig.~\ref{fig:MgvsCpMg_TMG} and Fig. S1) are consistent with previously reported results for MOCVD GaN:Mg layers. \cite{tokunaga1998growth,castiglia2011role,Tetsuo2018WideConcentrations,de2000influence}

In order to explore the limitations in Mg incorporation levels, additional GaN:Mg layers, M$_6$ - M$_9$, with the two highest Cp$_2$Mg/TMGa ratios of 0.335\% and 0.5\%  but with different V/III ratios, temperatures and growth rates were prepared. Samples M$_6$ and M$_7$ were grown with the same Cp$_2$Mg/TMGa ratio of $0.335\%$ as for sample M$_4$ but with  twice as high  V/III ratio of 1811 
and a reduced by half V/III ratio of 453 
, respectively. It can be noted that for constant CpMg/TMGa ratio  the increase of the V/III ratio leads to $1.5$ times lower Mg concentration, bringing it below the threshold for PID formation (for details on PID, see Sec. III B). When the V/III is decreased to 453, a significant increase in [Mg] to 7.0$\times{10}^{19}$ cm$^{-3}$ is observed (M$_7$). This is almost three times higher compared to the respective [Mg] in M$_4$ with a V/III ratio of 906. Note, that higher gas carrier flows of H$_2=25$~$\ell$/min and N$_2=12$~$\ell$/min (Table~\ref{tab:sample growth conditions}) were used for M$_7$ but their ratio was kept the same as for the case of samples M$_0$ - M$_6$. A detailed study on the effects of carrier gas flows and composition are reported elsewhere.\cite{AneliaAPL22} In a next step, a GaN:Mg with the same growth parameters as for M$_7$ but with a Cp$_2$Mg/TMGa of 0.5\% was grown (sample M$_8$) to investigate whether Mg incorporation can be further boosted by increasing the Cp2Mg/TMGa at these specific growth conditions. Interestingly, the [Mg] in M$_8$ is found to be $6.7\times{10}^{19}$ \si{{cm}^{-3}}, which is very similar to the [Mg] in M$_7$. 
This indicates that at low V/III ratios a saturation of Mg incorporation is already reached at  a Cp$_2$Mg/TMGa  of 0.335\%.  The lower V/III ratio is associated with a higher growth rate which can shift the balance from Mg atom substitutionally replacing a Ga atom towards remaining as an adatom providing potential explanation for the observed  Mg saturation. Finally, a GaN:Mg with a V/III ratio of 453, Cp$_2$Mg/TMGa ratio of 0.5\% but at a lower temperature of  \SI{1040}{\celsius} - M$_9$ was grown (the rest of the samples were all grown at \SI{1120}{\celsius}). This also led to a slightly different growth rate of 1.20 $\mu$m/h compared to M$_8$. 
The lower growth temperature resulted in high Mg incorporation level in the range of ${\sim1\times{10}^{20}}$ \si{{cm}^{-3}}.

Within the medium-to-high Mg concentration range up to 2.4$\times{10}^{19}$ cm$^{-3}$, [H] follows closely [Mg] being slightly lower than the expected 1:1 ratio (see Fig. \ref{fig:MgvsCpMg_TMG}, Table \ref{tab:table2} and Fig.~S2). Consequently, as-grown GaN:Mg exhibits $p$-type conductivity even before the annealing, as will be discussed later (see Sec. III C).  The H levels in all GaN:Mg layers with [Mg] above $2.4\times{10}^{19}$ \si{{cm}^{-3}} are   below ${\sim3\times{10}^{19}}$ \si{{cm}^{-3}}, being significantly lower than the respective [Mg] in agreement with earlier observations,  \cite{Tetsuo2018WideConcentrations,Kozodoy_2000}   (see Fig. \ref{fig:MgvsCpMg_TMG} and Fig.~S2).

\begin{figure}
\includegraphics[keepaspectratio=true,width=\linewidth, clip, trim=0cm 0cm 0cm 0cm ]{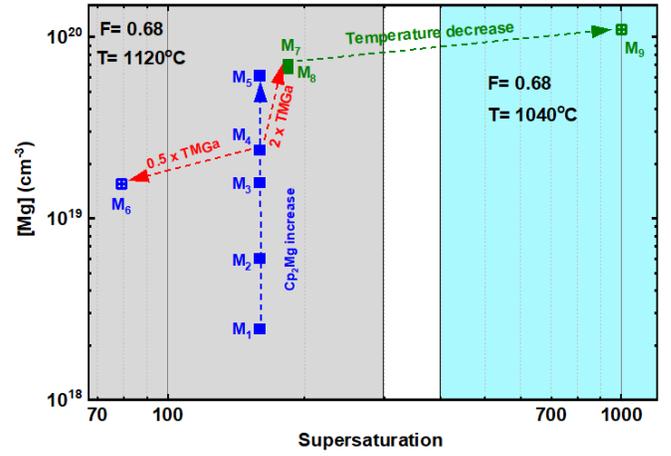}
\caption{Mg concentration ([Mg]) from Fig. \ref{fig:MgvsCpMg_TMG} as a function of Ga supersaturation in the as-grown GaN:Mg layers. The sample labels according to Table I are indicated in the figure. The same as in Fig. \ref{fig:MgvsCpMg_TMG}, blue symbols indicate samples grown with 19$\ell$ of H$_2$ and 9l N$_2$ while green symbols indicate  samples grown with 25$\ell$ of H$_2$ and 12$\ell$ N$_2$. The blue dashed arrow indicates the increase of Cp$_2$Mg/TMGa at constant supersaturation, the red dashed arrows indicate variation in supersaturation due to change of TMGa flow at constant Cp$_2$Mg/TMGa of 0.335\% and the green arrow indicates variation in supersaturation due to growth temperature at constant Cp$_2$Mg/TMGa of 0.5\% for specific sets of growth conditions. }
\label{fig:Mg vs Ga supersaturation}
\end{figure}

Our results indicate that the Mg incorporation in hot-wall MOCVD GaN is directly related to Cp$_2$Mg/TMGa ratio, more specifically to the Mg precursor flow in the reactor. In addition, V/III ratio, growth rate, growth temperature and pressure are impacting the growth process and hence incorporation of impurities and formation of native defects. It has been demonstrated that for MOCVD growth the concept of supersaturation, which measures the deviation from the thermodynamic equilibrium, is suitable to evaluate the impact of growth conditions in their complexity. \cite{MitaJAP08} Therefore, we present in Fig.~\ref{fig:Mg vs Ga supersaturation} [Mg] as a function of the Ga supersaturation. The latter was calculated following Ref. \onlinecite{MitaJAP08} and the respective growth parameters in Table I. Notably, the supersaturation has a strong effect on the incorporated [Mg] for samples with constant  Cp$_2$Mg/TMGa ratio. For instance at Cp$_2$Mg/TMGa of 0.335\%, the decrease of supersaturation from 160 (M$_4$) to 79 (M$_6$) results in 65\% decrease of [Mg] while increasing it to 184 (M$_7$) leads to a steep rise of [Mg] by 165\%. When the Ga supersaturation for constant Cp$_2$Mg/TMGa of 0.5\% increased from $184$ (M$_8$) to $1000$ (M$_9$) [Mg] is increased by 65\%. For the first case (M$_4$, M$_6$ and M$_7$), the increase of Ga supersaturation was achieved by varying the TMGa flow, and by decreasing the growth temperature for the latter case (M$_8$ - M$_9$ ). The ratio F (the ratio of the input mole fraction of H$_2$ to the total amount of the two carrier gases, H$_2$ and N$_2$),  which also strongly affects the supersaturation is the same for all samples. These observations show that the Mg incorporation in GaN is a multi-dimensional process and the trends related to different growth parameters e. g. V/III ratio, growth temperature, carrier gas composition etc., can be assessed by taking into account only the Cp$_2$Mg precursor flow and the Ga supersaturation. Clearly, Mg incorporation follows the same trend as Ga supersaturation i. e. the available Ga species in the reaction. Since Mg atoms compete with Ga atoms for incorporation at the same lattice site, at a constant Ga supersaturation the balance of  Mg/Ga atoms incorporation is shifted more towards the Mg side when the Cp$_2$Mg precursor flow i. e. the available Mg atoms in the reaction are increased. 


\begin{figure}
\includegraphics[keepaspectratio=true,width=\linewidth, clip, trim=0cm 0cm 0cm 0cm ]{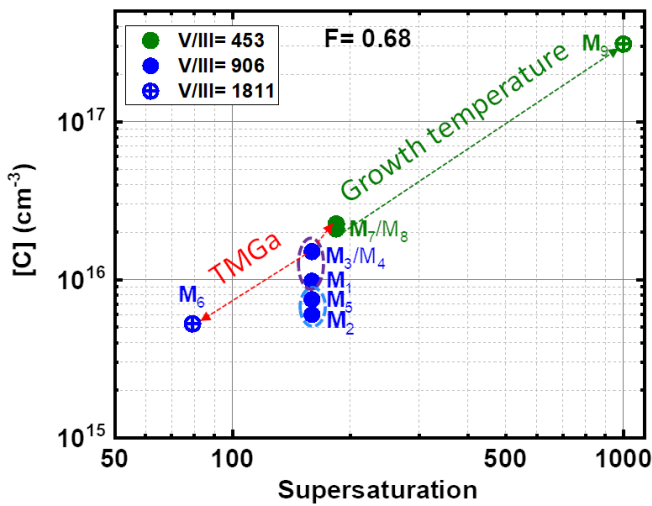}
\caption{Carbon concentration ([C]) in the as-grown GaN:Mg layers vs Ga supersaturation. The sample labels according to Table I are indicated in the figure. The same as in Fig. \ref{fig:MgvsCpMg_TMG}, blue symbol indicate samples grown with 19$\ell$ of H$_2$ and 9$\ell$ N$_2$ while green symbols indicate  samples grown with 25$\ell$ of H$_2$ and 12$\ell$ N$_2$.}
\label{fig:C vs Ga supersaturation}
\end{figure}


O and Si impurity levels in our GaN:Mg layers were found to be at the SIMS detection limit of (3-5)$\times{10}^{15}$ cm$^{-3}$ in all samples.  Hence only C from the unintentional impurities is further discussed in its potential role as a compensating donor and a trap affecting the free hole properties. The respective C concentrations in the GaN:Mg films are presented in Fig. \ref{fig:C vs Ga supersaturation} as a function of Ga supersaturation. An overall linear increase in [C] with increasing  Ga supersaturation is evident independently of whether the change in supersaturation is achieved via varying  V/III ratio or growth temperature. The increase in [C] at constant supersaturation could be associated with the increase of the Cp$_2$Mg/TMGa ratio. \footnote{Samples M$_2$ and M$_5$ were grown with fresh satellite, which could explain the slight variations in comparison to samples M$_1$,  M$_3$ and M$_4$. We note that this has no baring on the reported general trends and conclusions.}

\subsection{Effect of Mg on the extended defects and surface morphology}

As can be seen from Table \ref{tab:sample properties}, a small variation of the screw-type dislocations density upon increasing the Mg concentration is observed in the as-grown GaN:Mg layers. The $N\mathrm{_{S}}$ remains very similar to the value found in the undoped GaN layer. On the other hand, $N_\mathrm{E}$ increases from (5-6)$\times{10}^8$ cm$^{-2}$ to (7-9)$\times{10}^8$ cm$^{-2}$ for [Mg] $\geq 2.4\times{10}^{19}$ cm$^{-3}$. For a comparison, in Mg doped InN layers, the edge dislocation density remains similar with increasing Mg, while screw dislocation density increases accompanied by a polarity inversion. \cite{XieJAP14}

\begin{figure}
\includegraphics[keepaspectratio=true,width=\linewidth, clip, trim=0cm 0cm 0cm 0cm ]{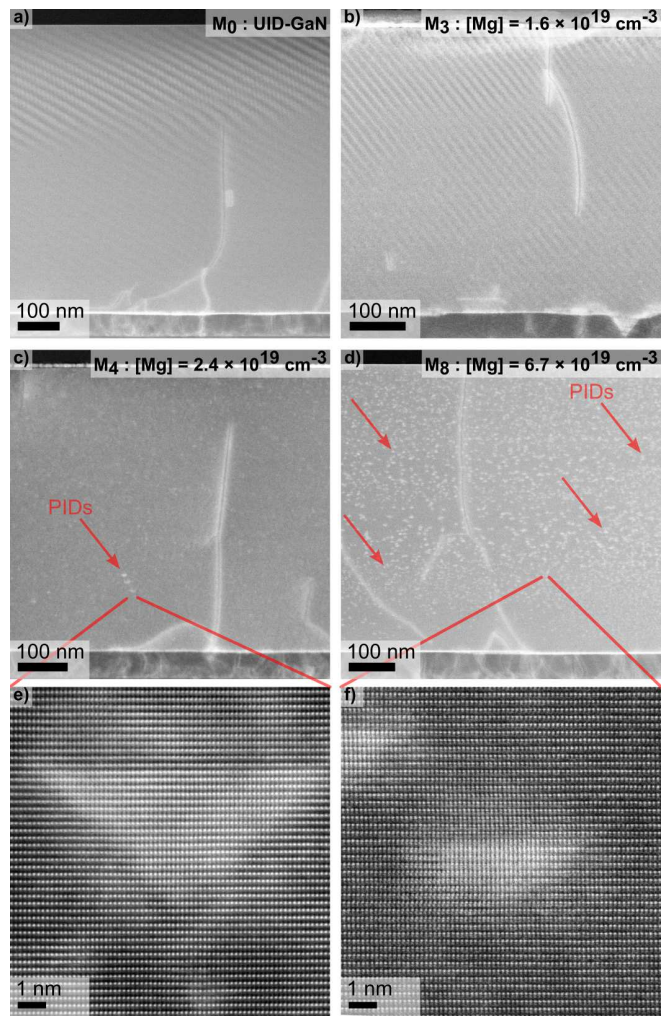}
\caption{STEM images of as-grown GaN:Mg layers with different [Mg]: (a) M$_0$ - UID, (b) M$_3$ - $1.6\times{10}^{19}$ cm$^{-3}$, (c) M$_4$ - $2.4\times{10}^{19}$ cm$^{-3}$, and (d) M$_8$ - $6.7\times{10}^{19}$ cm$^{-3}$. Arrows in (c) and (d) highlight some of the polarity inversion domains (PIDs). Higher magnification images of PIDs for M$_4$  and M$_8$ are shown in (e) and (f), respectively (both in the $[1\bar{1}00]$ projection).}
\label{TEM}
\end{figure}

\begin{figure*}
\includegraphics[width=0.99\textwidth]{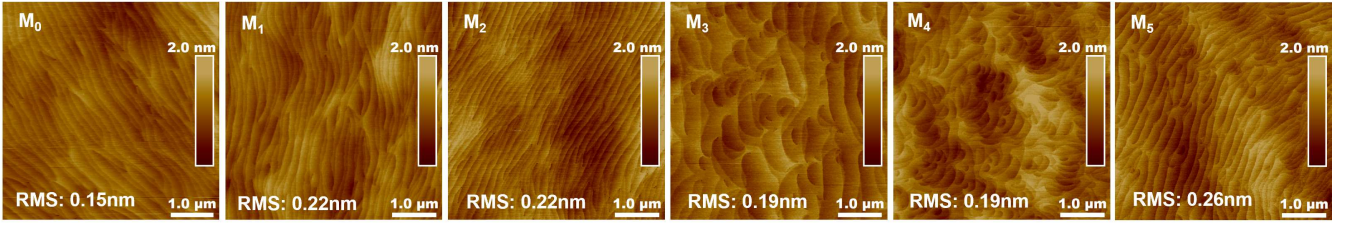}
 \caption{$5\times{5}$ \si{\mu m^{2}} AFM images of as-grown GaN:Mg layers  with different [Mg]: M$_0$ - UID, M$_1$ - $2.5\times{10}^{18}$ \si{cm^{-3}}, M$_2$ - $6.0\times{10}^{18}$ \si{cm^{-3}}, M$_3$ - $1.6\times{10}^{19}$ \si{cm^{-3}}, M$_4$ - $2.4\times{10}^{19}$ \si{cm^{-3}} and M$_5$ - $6.1\times{10}^{19}$ \si{cm^{-3}}. For further details on the samples see Table \ref{tab:sample growth conditions} and \ref{tab:sample properties}.}
 \label{fig:AFM}
\end{figure*}

STEM analysis was performed to provide further insight into the microstructure of the GaN:Mg layers. The results shown in Fig. \,\ref{TEM} verify that extended defects are not generated in layers doped with Mg up to $1.6\times{10}^{19}$ \si{cm^{-3}} (Fig. \,\ref{TEM} (a) and (b)). For [Mg]\,=\,$2.4\times{10}^{19}$ \si{cm^{-3}} (sample M$_4$) sporadic PID defects are present (Fig. \,\ref{TEM} (c) and (e)). Incorporating more Mg (sample M$_8$ with [Mg]\,=\,$6.7\times{10}^{19}$) leads to a notable increase in PID density (Fig. \,\ref{TEM} (d)). The average width of the PIDs in the latter case is estimated to be $5.4\pm0.9$ nm (Fig. \,\ref{TEM} (d)), in range with previously reported values.  \cite{Tetsuo2018WideConcentrations,Narita_2020} Polarity inversion inside the pyramids is confirmed by annular bright-field (ABF) STEM (see Fig. S3 in the supplementary material). STEM analysis further revealed segregation of Mg at the PID ($0001$) facet (Fig. S3) in concordance with earlier observations. \cite{Tetsuo2018WideConcentrations,Narita_2020} Additionally, our comprehensive EELS analysis reveals that Mg atoms are segregated not only at the ($0001$) planes of the PIDs, but they are also found at their inclined facets (Further details on the EELS analysis will be reported elsewhere).

A crude estimate of the PID density results in a range between $1.9$ and $8.9$ $\times{10}^{16}$ cm$^{-3}$ by sampling different regions of M$_8$ (Fig. \,\ref{TEM} (d)). The extended range originates from a thickness measurement by EELS acquired in an adjacent region. An estimate of the  number of Mg atoms associated with the PIDs can be obtained following Narita \emph{et al.}, \cite{Narita_2020,Tetsuo2018WideConcentrations} for calculating the Mg atoms bound to the top facet and adding the herein observed segregation of Mg to the side facets. Accordingly, the [Mg] bound to the PIDs in M$_8$ is estimated to be from 1.6$\times{10}^{19}$ cm$^{-3}$ to 7.7$\times{10}^{19}$ cm$^{-3}$. We expect the actual value to be on the higher end of the range, since the PID density measurement was performed close to an edge where the measured thickness is on the lower end of our estimate. In comparison, the difference between [Mg] and [H] in this sample as measured by SIMS amounts to 5.3$\times{10}^{19}$ cm$^{-3}$ (Table \ref{tab:sample properties} and Fig.\,S2). It is therefore inferred that the difference between [Mg] and [H] at doping levels above 2.4$\times{10}^{19}$ cm$^{-3}$ can be accounted for by Mg segregated at all the interfaces of the PIDs.

\begin{figure}
\includegraphics[width=0.5\textwidth]{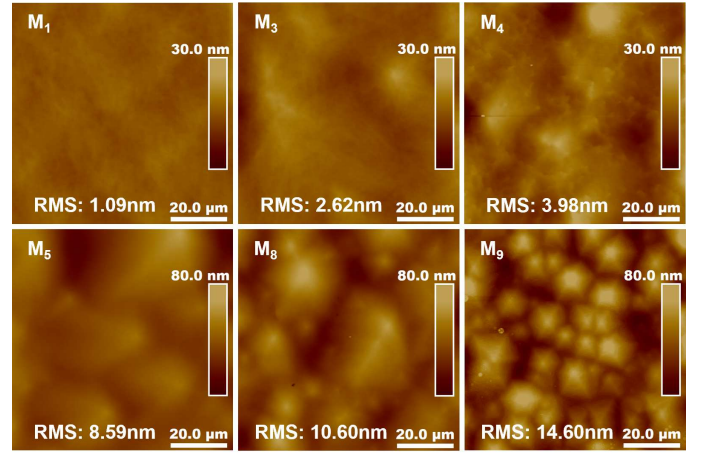}
 \caption{Large-area $80\times{80}$ \si{\mu m^{2}} AFM images of GaN layers doped with different Mg levels: M$_1$ - $2.5\times{10}^{18}$ \si{cm^{-3}}, M$_3$ - $1.6\times{10}^{19}$ \si{cm^{-3}}, M$_4$ - $2.4\times{10}^{19}$ \si{cm^{-3}}, M$_5$ - $6.10\times{10}^{19}$ \si{cm^{-3}}, M$_8$ - $6.7\times{10}^{19}$ \si{cm^{-3}} and M$_9$ - $1.1\times{10}^{20}$ \si{cm^{-3}}. Note the different scale bar for images on the top and bottom rows. For further details on the samples see Table \ref{tab:sample growth conditions} and \ref{tab:sample properties}.}
 \label{fig:AFMlarge}
\end{figure}

The presence of PIDs is also expected to affect the surface morphology, in particular for the high-defect-density layers with [Mg] \,$\geq$\, $2.4\times{10}^{19}$ cm$^{-3}$. The UID GaN and GaN:Mg layers M$_0$ - M$_5$ grown with the optimal growth conditions exhibit step-flow growth mode and the terrace size is similar for the undoped and doped layers up to [Mg] of $6.1\times{10}^{19}$ cm$^{-3}$ as can be seen from the 5$\times$5 $\mu$m$^{2}$ micrographs in Fig.~\ref{fig:AFM}. The root-mean-square roughness (RMS) for M$_0$ - M$_5$ is in the range of 0.15 - 0.26 nm.  However, on a larger-scale, changes in morphology could be observed already at moderate Mg concentrations. Figure~\ref{fig:AFMlarge} shows representative 80$\times$80 $\mu$m$^{2}$ micrographs of selected samples. It is seen that already for Mg doping of $1.6\times{10}^{19}$ cm$^{-3}$ (M$_3$) there is a slight increase of RMS as compared to M$_1$. The RMS of sample M$_4$, for which sporadic PIDs start to appear, is further increased by a factor of two (Fig. \ref{fig:AFMlarge}). With further increase of Mg concentration one can observe well defined hexagonal hillocks (Fig. \,\ref{TEM} (d)). Their density and height is increasing with increasing [Mg], which is manifested in higher RMS surface roughness. This surface deterioration can be correlated with the enhanced PID density that takes place with increasing [Mg]. Previously, it was shown that for [Mg] \,$\geq 1.0\times{10}^{20}$ \si{cm^{-3}} the formation of hexagonal hillocks and platelets can be associated with a polarity inversion to N-polar GaN.  \cite{RomanoAPL99,Tavernier2004TheIn,N-polarMgdopedGaN-APL07} Our results indicate that this is a gradual process that starts already at [Mg] \,=\,2.4$\times{10}^{19}$ \si{cm^{-3}} and is associated with the PID formation in the overall Ga-polar GaN. Note that this is not accompanied by any noticeable increase  of screw dislocation densities as earlier suggested for the case of full polarity inversion in GaN \cite{RomanoAPL99} and experimentally observed for InN.  \cite{XieJAP14} For example, samples M$_3$ and M$_8$ have very similar $N_S$ (Table \ref{tab:sample properties}) while their PID density differ by orders of magnitude. 
 
\subsection{Electrical and free-hole properties}

\begin{figure}
\includegraphics[keepaspectratio=true,width=\linewidth, clip, trim=0cm 0cm 0cm 0cm ]{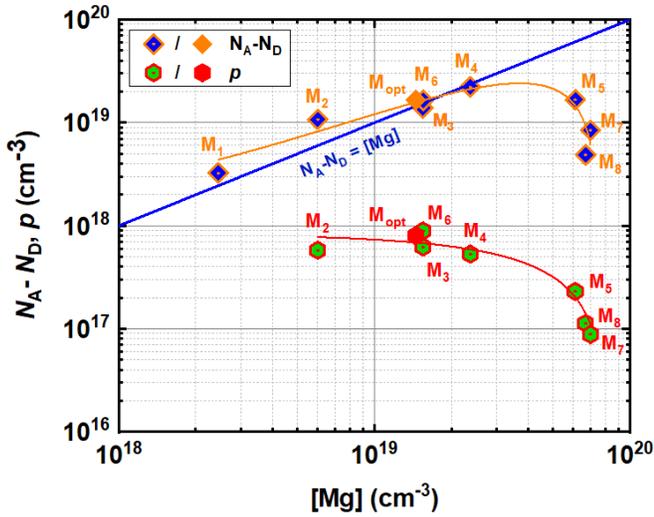}
\caption{Net acceptor concentration $N\mathrm{_A}-N\mathrm{_D}$ (from C-V measurements) and free-hole concentrations (from Hall) in the annealed $p$-GaN as a function of [Mg]. The red and orange curves are guide to the eye. The blue solid line corresponds to $N\mathrm{_A}-N\mathrm{_D}$ = [Mg].}
\label{fig:N_A-N_D and p vs Mg}
\end{figure}

\begin{figure}
\includegraphics[keepaspectratio=true,width=\linewidth, clip, trim=0cm 0cm 0cm 0cm ]{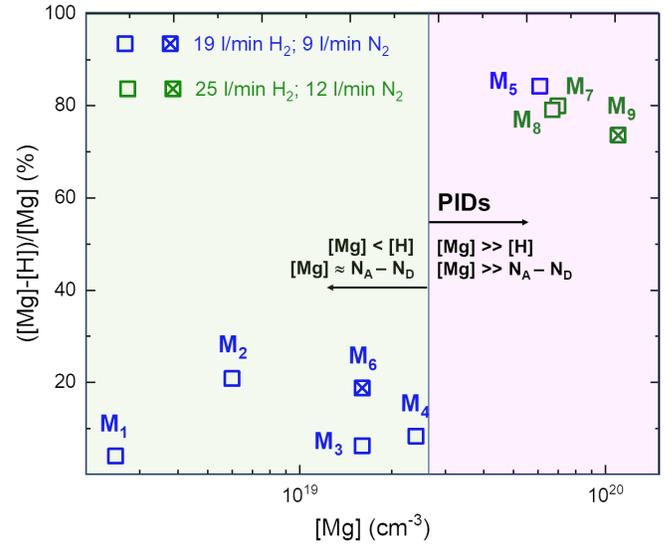}
\caption{Fraction of non-passivated Mg, ([Mg] - [H])/[Mg] in the as-grown GaN:Mg layers as a function of Mg doping.}
\label{fig:inactiveMg}
\end{figure}

Figure~\ref{fig:N_A-N_D and p vs Mg} shows the net acceptor concentration $N\mathrm{_A}-N\mathrm{_D}$ and the free-hole concentrations in the annealed samples. The relative difference between Mg and H concentrations with respect to the total Mg concentration, ([Mg] - [H])/[Mg] in the as-grown GaN:Mg is plotted as a function of [Mg] in Fig. \,\ref{fig:inactiveMg}. Assuming that all H binds to the available Mg$_{\mathrm{Ga}}$,\cite{WalleAPL96} this fraction represents the total amount of Mg atoms not bound to H and it may include: $i$) Mg$_{\mathrm{Ga}}$ acceptors not passivated by H and/or $ii$) Mg incorporated at a different site in the GaN crystal lattice, e. g., Mg interstitial, Mg segregated on the PID facets etc. Two distinct behaviors of ($N\mathrm{_A}-N\mathrm{_D}$)/$p$ and  ([Mg] - [H])/[Mg] can be discerned below and above  [Mg] = 2.4$\times{10}^{19}$ cm$^{-3}$, respectively.

 Up to 2.4$\times{10}^{19}$ cm$^{-3}$ the $N\mathrm{_A}-N\mathrm{_D}$  follows closely the [Mg]  indicating that Mg is incorporated as an acceptor within this doping range. 
A certain fraction between 4\% and 21\% of [Mg] remains non-passivated by H (Fig. \ref{fig:inactiveMg} and Table~\ref{tab:sample properties}). Since all Mg atoms are incorporated as acceptors (Fig. \ref{fig:N_A-N_D and p vs Mg}), this result implies that  [Mg] - [H])/[Mg] represents the fraction of Mg atoms of type $i$), i. e. the isolated Mg$_\mathrm{Ga}$ acceptor not passivated by H. \cite{WallePRL12}  As a result, as-grown GaN:Mg manifests a net $p$-type conductivity with free hole concentration in the low to high ${10}^{16}$ cm$^{-3}$ range (samples M$_2$, M$_3$ and M$_6$, see Table~\ref{tab:sample properties}). These findings are very interesting since they are contrasted with previous reports, where [H] is equal or larger than [Mg] \cite{Tetsuo2018WideConcentrations,castiglia2011role} and indicate the expanded opportunities offered by the hot-wall MOCVD concept.  The higher the non-passivated Mg fraction, the higher the free hole concentration in the as-grown GaN:Mg for this doping range. Details on exploring carrier gas composition for increasing the non-passivated Mg acceptor and free-hole concentrations will be presented elsewere.\cite{AneliaAPL22} It is instructive to compare samples M$_3$ and M$_6$ with the same [Mg] but significantly different levels of non-passivated Mg of 1$\times{10}^{18}$ cm$^{-3}$ and 3$\times{10}^{18}$ cm$^{-3}$, respectively. The comparison reveals that M$_6$ having three times larger fraction of non-passivated Mg exhibits four times higher free hole concentration before annealing with regard to M$_3$. Assuming ionization energy of 180-200 meV corresponding to N$_\mathrm{A}$ of 1$\times{10}^{18}$ cm$^{-3}$ and 3$\times{10}^{18}$ according to Ref. [7], the maximum free-hole concentrations in the as-grown M$_3$ and M$_6$ layers are expected to be in the mid and high ${10}^{16}$ cm$^{-3}$ range, respectively. 
The $p_0$ values measured for M$_3$ and M$_6$ are  lower indicating certain degree of compensation. We recall that M$_6$ is grown at higher V/III ratio as compared to M$_3$, i. e., at lower supersaturation (Table \ref{tab:Sample growth conditions}) and exhibits significantly lower levels of C (Table~\ref{tab:sample properties} and Fig. \,\ref{fig:C vs Ga supersaturation}). Under the N-rich conditions, typically employed in MOCVD growth, the C$_\mathrm{Ga}$ acting as a donor  has the lowest formation energy in $p$-type GaN. \cite{lyons2014effects} Although [C] is at least two orders of magnitude lower than the non-passivated isolated Mg acceptors ([Mg] - [H]), it is close to the observed free-hole concentrations in the as-grown layers and will interfere negatively with the $p$-type conductivity. This can explain the observed lower free-hole concentration before annealing in M$_3$ as compared to M$_6$. After annealing the free-hole concentration in M$_6$ is also higher as compared with M$_3$, correlating with $N_\mathrm{A} - N_\mathrm{D}$. In this case, $N_\mathrm{A} - N_\mathrm{D}$ is dominated by the thermally activated Mg acceptors resulting from the dissociation of H from the Mg-H complexes in the as-grown samples.
 

The increase of Mg concentration above 2.4$\times{10}^{19}$ cm$^{-3}$ leads to a decrease of $N\mathrm{_A}-N\mathrm{_D}$ which becomes significantly lower than [Mg] (Fig. \,\ref{fig:N_A-N_D and p vs Mg}), consistent with previous reports.\cite{castiglia2011role, Tetsuo2018WideConcentrations} The C-V measurements show that 72\% to 92\% of the incorporated Mg atoms are not electrically active. This doping range corresponds to [H] $\ll$ [Mg] in the as-grown samples as can be seen from Figs.\,\ref{fig:MgvsCpMg_TMG} and S2 and the respective fraction of Mg not-bound to H is in the range of 74\% to 84\% with no distinct dependence on [Mg] (Fig. \,\ref{fig:inactiveMg}). These results imply that the number of Mg atoms not binding to H in the as-grown layers could account to a large extent for the difference between the net acceptor and Mg concentrations in the respective samples after annealing. In other words, the fraction ([Mg] - [H])/[Mg] corresponds to scenario $ii$) in which Mg incorporates at non-acceptor sites in the GaN crystal lattice. 

 This is reflected in a significant reduction of free-hole concentration in the annealed samples (Fig. \,\ref{fig:N_A-N_D and p vs Mg}). Note that the lower $N\mathrm{_A}-N\mathrm{_D}$  (Fig. \ref{fig:N_A-N_D and p vs Mg}) the lower the free hole concentration in the as-grown (Table~\ref{tab:sample properties}) and annealed  (Fig. \ref{fig:N_A-N_D and p vs Mg}) samples. Our STEM analysis shows that the observed large fraction of electrically inactive Mg could potentially be explained by the estimated concentration of Mg segregated at the PIDs. However, the fraction of Mg not bound to H is very similar for samples M$_5$, M$_7$ and M$_8$ (Fig. \ref{fig:inactiveMg}) while their $N\mathrm{_A}-N\mathrm{_D}$ differs significantly (Fig. \ref{fig:N_A-N_D and p vs Mg}). For example, sample M$_5$ with inactive Mg fraction of 84\% has more than twice as high net acceptor concentration than M$_8$ with 74\% of inactive [Mg] non-bound to H, while the opposite could be expected. In fact, in M$_5$, the concentration of Mg-H complexes before annealing is  9.6$\times{10}^{18}$ cm$^{-3}$ (assuming all available H binds to Mg), which compares well with the $N\mathrm{_A}-N\mathrm{_D}$  of 1.7$\times{10}^{19}$ cm$^{-3}$ after annealing.  This suggests that no significant concentrations of compensating donors, most notably the V$_\mathrm{N}$ and its complexes with Mg, are generated. In samples M$_7$ and M$_8$, the  Mg-H levels before annealing are slightly higher as compared to M$_5$, while $N\mathrm{_A}-N\mathrm{_D}$ are significantly lower, i. e. (5-8)$\times{10}^{18}$ cm$^{-3}$, indicating that compensating defects are present in large densities. Both M$_7$ and M$_8$ are grown at lower V/III ratio, which can explain an enhanced formation of V$_\mathrm{N}$ in this case.  Our findings strongly suggest that compensating donors, likely V$_\mathrm{N}$ and its complexes with Mg are generated for highly Mg doped GaN grown at higher supersaturation and which play a significant role for the observed reduction of free-hole concentration (Fig. \ref{fig:N_A-N_D and p vs Mg}).  The increase of [Mg] above 1$\times{10}^{20}$, does not alter the fraction of non-passivated electrically inactive Mg. In this case, however, the high [C] levels of 3.1$\times{10}^{17}$ cm$^{-3}$, due to the significantly lower growth temperature, lead to additional compensation and the GaN layer shows semi-insulating behavior.

It has been shown that by growing GaN:Mg on GaN substrates, with significantly lower density of dislocations, the free hole concentration in the homoepitaxial layers can be significantly enhanced.  \cite{Koide_APL19} To reduce the dislocation density in our heteroepitaxial case we grew M$_\mathrm{opt}$ consisting of a 550-nm-thick GaN:Mg layer on 1-$\mu$m-thick undoped GaN buffer layer using the same growth conditions as for sample M$_3$.  Both screw- and edge-type dislocation densities of M$_\mathrm{opt}$ show the lowest values of $1.6\times{10}^{7}$ cm$^{-2}$ and $4.3\times{10}^{8}$ cm$^{-2}$, respectively, among all samples (Table\,\ref{tab:sample properties}).  $N_\mathrm{S}$ and  $N_\mathrm{E}$ in the GaN:Mg are likely to be even lower since the measured dislocation densities are averaged over the entire layer thickness.   The free-hole concentration in the as-grown M$_\mathrm{opt}$ is increased more than twice as compared to the respective value in M$_3$. If similar approach is employed for lower Mg doping levels, e. g. such as in M$_2$, free-hole concentrations in the low to mid ${10}^{17}$ cm$^{-3}$ range can be potentially achieved without the need of thermal activation. The free-hole concentration of 8.4$\times{10}^{17}$ cm$^{-3}$ and the resistivity of $0.77$ $\Omega$.cm in the annealed sample are among the best results reported in literature. \cite{Narita_2020,Koide_APL19} These results demonstrate that the hot-wall MOCVD capabilities can be expanded to deliver state-of-the art $p$-type GaN enabling development of power diodes and vertical transistors, and normally-off HEMTs, for example.

\section{Conclusions}

We have explored hot-wall MOCVD for the growth of Mg doped GaN. A detailed study of growth and doping conditions impacts on the incorporation of Mg, H and C is presented and the findings are discussed in terms of Ga supersaturation. Our results clearly indicate that Ga supersaturation is a universal parameter that can be conveniently used for the optimization of Mg incorporation  and C reduction in MOCVD of GaN instead of multiple growth parameters independently used for the same purpose. 

In the low doping range ([Mg]\,$\leq$\,2.4$\times{10}^{19}$ cm$^{-3}$), a large fraction (up to 21\%) of active Mg acceptors (not passivated by H)  is evidenced in as-grown material. Hence, hot-wall MOCVD provides an intriguing opportunity to exploit the technique for delivering lightly p-type doped material for e. g. vertical MOSFETs, without the need of post-growth annealing. In the annealed samples, $N_\mathrm{A} - N_\mathrm{D}$ is found to closely follow [Mg] indicating that all Mg atoms are incorporated as acceptors, where 96\% are activated by thermal dissociation of H during post-growth annealing. 

In the high doping range above 2.4$\times{10}^{19}$ cm$^{-3}$, increased Mg leads to the generation of PIDs. It is shown that this can be correlated with the formation of hexagonal hillocks, the density and height of which increase with increasing [Mg]. This gradual surface morphology deterioration can be clearly observed in large-scale AFM imaging while on a smaller scale the morphology and surface roughness remain virtually unchanged. STEM analysis reveals segregation of Mg at the PIDs, sufficient to account for the observed amount of electrically inactive Mg not passivated by H. However, generation of V$_\mathrm{N}$ and its complexes with Mg$_\mathrm{Ga}$ need to be invoked to explain the considerably lower free hole and net acceptor concentrations measured in the highly doped GaN:Mg grown at high supersaturation.   

Under optimized growth conditions high-quality GaN:Mg with resistivity of 0.77 $\Omega$.cm and free-hole concentration of 8.4$\times{10}^{17}$ cm$^{-3}$ corresponding to Mg acceptor activation of 5.25\% has been demonstrated. These results and the established comprehensive picture of GaN:Mg growth via hot-wall MOCVD substantiate the expanded capabilities of the technique to deliver state-of-the-art $p$-type GaN for the development of power diodes and transistors.

\section*{Supplementary Material}

See the supplementary material for additional details about Mg and H incorporation in the studied layers as well as on the structure of the observed pyramidal defects as revealed by STEM.

\section*{Acknowledgments}
This work is performed within the framework of the competence center for III-Nitride technology, C3NiT - Janz\'en supported by the Swedish Governmental
Agency for Innovation Systems (VINNOVA) under
the Competence Center Program Grant No. 2016-05190, Link\"oping University, Chalmers University of technology, Ericsson, Epiluvac, FMV, Gotmic, Hexagem, Hitachi Energy, On Semiconductor, Saab, SweGaN, and UMS. We further acknowledge support from the the Swedish Research Council VR under Award No. 2016-00889, Swedish Foundation for Strategic Research under Grants No. RIF14-055, RIF14-0074 and No. EM16-0024, and the Swedish Government Strategic Research Area in Materials Science on Functional Materials at Link\"oping University, Faculty Grant SFO Mat LiU No. 2009-00971. The KAW Foundation is also acknowledged for support of the Link\"oping Electron Microscopy Laboratory. 

\section*{Author Declaration}

The authors declare no conflict of interest.

\section*{Data Availability Statement}

The data that support the findings of this study are available
within the article and its supplementary material.

\section*{References}

\bibliography{library_1}

\begin{thebibliography}{37}%
\makeatletter
\providecommand \@ifxundefined [1]{%
 \@ifx{#1\undefined}
}%
\providecommand \@ifnum [1]{%
 \ifnum #1\expandafter \@firstoftwo
 \else \expandafter \@secondoftwo
 \fi
}%
\providecommand \@ifx [1]{%
 \ifx #1\expandafter \@firstoftwo
 \else \expandafter \@secondoftwo
 \fi
}%
\providecommand \natexlab [1]{#1}%
\providecommand \enquote  [1]{``#1''}%
\providecommand \bibnamefont  [1]{#1}%
\providecommand \bibfnamefont [1]{#1}%
\providecommand \citenamefont [1]{#1}%
\providecommand \href@noop [0]{\@secondoftwo}%
\providecommand \href [0]{\begingroup \@sanitize@url \@href}%
\providecommand \@href[1]{\@@startlink{#1}\@@href}%
\providecommand \@@href[1]{\endgroup#1\@@endlink}%
\providecommand \@sanitize@url [0]{\catcode `\\12\catcode `\$12\catcode
  `\&12\catcode `\#12\catcode `\^12\catcode `\_12\catcode `\%12\relax}%
\providecommand \@@startlink[1]{}%
\providecommand \@@endlink[0]{}%
\providecommand \url  [0]{\begingroup\@sanitize@url \@url }%
\providecommand \@url [1]{\endgroup\@href {#1}{\urlprefix }}%
\providecommand \urlprefix  [0]{URL }%
\providecommand \Eprint [0]{\href }%
\providecommand \doibase [0]{http://dx.doi.org/}%
\providecommand \selectlanguage [0]{\@gobble}%
\providecommand \bibinfo  [0]{\@secondoftwo}%
\providecommand \bibfield  [0]{\@secondoftwo}%
\providecommand \translation [1]{[#1]}%
\providecommand \BibitemOpen [0]{}%
\providecommand \bibitemStop [0]{}%
\providecommand \bibitemNoStop [0]{.\EOS\space}%
\providecommand \EOS [0]{\spacefactor3000\relax}%
\providecommand \BibitemShut  [1]{\csname bibitem#1\endcsname}%
\let\auto@bib@innerbib\@empty
\bibitem [{\citenamefont {Kozodoy}\ \emph {et~al.}(2000)\citenamefont
  {Kozodoy}, \citenamefont {Xing}, \citenamefont {DenBaars}, \citenamefont
  {Mishra}, \citenamefont {Saxler}, \citenamefont {Perrin}, \citenamefont
  {Elhamri},\ and\ \citenamefont {Mitchel}}]{Kozodoy_2000}%
  \BibitemOpen
  \bibfield  {author} {\bibinfo {author} {\bibfnamefont {P.}~\bibnamefont
  {Kozodoy}}, \bibinfo {author} {\bibfnamefont {H.}~\bibnamefont {Xing}},
  \bibinfo {author} {\bibfnamefont {S.~P.}\ \bibnamefont {DenBaars}}, \bibinfo
  {author} {\bibfnamefont {U.~K.}\ \bibnamefont {Mishra}}, \bibinfo {author}
  {\bibfnamefont {A.}~\bibnamefont {Saxler}}, \bibinfo {author} {\bibfnamefont
  {R.}~\bibnamefont {Perrin}}, \bibinfo {author} {\bibfnamefont
  {S.}~\bibnamefont {Elhamri}}, \ and\ \bibinfo {author} {\bibfnamefont
  {W.}~\bibnamefont {Mitchel}},\ }\href@noop {} {\bibfield  {journal} {\bibinfo
   {journal} {J. Appl. Phys.}\ }\textbf {\bibinfo {volume} {87}},\ \bibinfo
  {pages} {1832} (\bibinfo {year} {2000})}\BibitemShut {NoStop}%
\bibitem [{\citenamefont {Narita}\ \emph {et~al.}(2018)\citenamefont {Narita},
  \citenamefont {Ikarashi}, \citenamefont {Tomita}, \citenamefont {Kataoka},\
  and\ \citenamefont {Kachi}}]{Tetsuo2018WideConcentrations}%
  \BibitemOpen
  \bibfield  {author} {\bibinfo {author} {\bibfnamefont {T.}~\bibnamefont
  {Narita}}, \bibinfo {author} {\bibfnamefont {N.}~\bibnamefont {Ikarashi}},
  \bibinfo {author} {\bibfnamefont {K.}~\bibnamefont {Tomita}}, \bibinfo
  {author} {\bibfnamefont {K.}~\bibnamefont {Kataoka}}, \ and\ \bibinfo
  {author} {\bibfnamefont {T.}~\bibnamefont {Kachi}},\ }\href {\doibase
  10.1063/1.5045257} {\bibfield  {journal} {\bibinfo  {journal} {J. Appl.
  Phys.}\ }\textbf {\bibinfo {volume} {124}},\ \bibinfo {pages} {165706}
  (\bibinfo {year} {2018})}\BibitemShut {NoStop}%
\bibitem [{\citenamefont {Narita}\ \emph {et~al.}(2020)\citenamefont {Narita},
  \citenamefont {Tomita}, \citenamefont {Kataoka}, \citenamefont {Tokuda},
  \citenamefont {Kogiso}, \citenamefont {Yoshida}, \citenamefont {Ikarashi},
  \citenamefont {Iwata}, \citenamefont {Nagao}, \citenamefont {Sawada},
  \citenamefont {Horita}, \citenamefont {Suda},\ and\ \citenamefont
  {Kachi}}]{Narita_2020}%
  \BibitemOpen
  \bibfield  {author} {\bibinfo {author} {\bibfnamefont {T.}~\bibnamefont
  {Narita}}, \bibinfo {author} {\bibfnamefont {K.}~\bibnamefont {Tomita}},
  \bibinfo {author} {\bibfnamefont {K.}~\bibnamefont {Kataoka}}, \bibinfo
  {author} {\bibfnamefont {Y.}~\bibnamefont {Tokuda}}, \bibinfo {author}
  {\bibfnamefont {T.}~\bibnamefont {Kogiso}}, \bibinfo {author} {\bibfnamefont
  {H.}~\bibnamefont {Yoshida}}, \bibinfo {author} {\bibfnamefont
  {N.}~\bibnamefont {Ikarashi}}, \bibinfo {author} {\bibfnamefont
  {K.}~\bibnamefont {Iwata}}, \bibinfo {author} {\bibfnamefont
  {M.}~\bibnamefont {Nagao}}, \bibinfo {author} {\bibfnamefont
  {N.}~\bibnamefont {Sawada}}, \bibinfo {author} {\bibfnamefont
  {M.}~\bibnamefont {Horita}}, \bibinfo {author} {\bibfnamefont
  {J.}~\bibnamefont {Suda}}, \ and\ \bibinfo {author} {\bibfnamefont
  {T.}~\bibnamefont {Kachi}},\ }\href@noop {} {\bibfield  {journal} {\bibinfo
  {journal} {Jpn. J. Appl. Phys.}\ }\textbf {\bibinfo {volume} {59}},\ \bibinfo
  {pages} {SA0804} (\bibinfo {year} {2020})}\BibitemShut {NoStop}%
\bibitem [{\citenamefont {G{\"o}tz}\ \emph {et~al.}(1995)\citenamefont
  {G{\"o}tz}, \citenamefont {Johnson}, \citenamefont {Walker}, \citenamefont
  {Bour}, \citenamefont {Amano},\ and\ \citenamefont {Akasaki}}]{GotzAPL95}%
  \BibitemOpen
  \bibfield  {author} {\bibinfo {author} {\bibfnamefont {W.}~\bibnamefont
  {G{\"o}tz}}, \bibinfo {author} {\bibfnamefont {N.~M.}\ \bibnamefont
  {Johnson}}, \bibinfo {author} {\bibfnamefont {J.}~\bibnamefont {Walker}},
  \bibinfo {author} {\bibfnamefont {D.~P.}\ \bibnamefont {Bour}}, \bibinfo
  {author} {\bibfnamefont {H.}~\bibnamefont {Amano}}, \ and\ \bibinfo {author}
  {\bibfnamefont {I.}~\bibnamefont {Akasaki}},\ }\href@noop {} {\bibfield
  {journal} {\bibinfo  {journal} {Appl. Phys. Lett.}\ }\textbf {\bibinfo
  {volume} {67}},\ \bibinfo {pages} {2666} (\bibinfo {year}
  {1995})}\BibitemShut {NoStop}%
\bibitem [{\citenamefont {Amano}\ \emph {et~al.}(1989)\citenamefont {Amano},
  \citenamefont {Kito}, \citenamefont {Hiramatsu},\ and\ \citenamefont
  {Akasaki}}]{amano1989p}%
  \BibitemOpen
  \bibfield  {author} {\bibinfo {author} {\bibfnamefont {H.}~\bibnamefont
  {Amano}}, \bibinfo {author} {\bibfnamefont {M.}~\bibnamefont {Kito}},
  \bibinfo {author} {\bibfnamefont {K.}~\bibnamefont {Hiramatsu}}, \ and\
  \bibinfo {author} {\bibfnamefont {I.}~\bibnamefont {Akasaki}},\ }\href@noop
  {} {\bibfield  {journal} {\bibinfo  {journal} {Jpn. J. Appl. Phys.}\ }\textbf
  {\bibinfo {volume} {28}},\ \bibinfo {pages} {L2112} (\bibinfo {year}
  {1989})}\BibitemShut {NoStop}%
\bibitem [{\citenamefont {Nakamura}\ \emph {et~al.}(1992)\citenamefont
  {Nakamura}, \citenamefont {Mukai}, \citenamefont {Senoh},\ and\ \citenamefont
  {Iwasa}}]{Nakamura_1992}%
  \BibitemOpen
  \bibfield  {author} {\bibinfo {author} {\bibfnamefont {S.}~\bibnamefont
  {Nakamura}}, \bibinfo {author} {\bibfnamefont {T.}~\bibnamefont {Mukai}},
  \bibinfo {author} {\bibfnamefont {M.}~\bibnamefont {Senoh}}, \ and\ \bibinfo
  {author} {\bibfnamefont {N.}~\bibnamefont {Iwasa}},\ }\href {\doibase
  10.1143/jjap.31.l139} {\bibfield  {journal} {\bibinfo  {journal} {Jpn. J.
  Appl. Phys.}\ }\textbf {\bibinfo {volume} {31}},\ \bibinfo {pages} {L139}
  (\bibinfo {year} {1992})}\BibitemShut {NoStop}%
\bibitem [{\citenamefont {Brochen}\ \emph {et~al.}(2013)\citenamefont
  {Brochen}, \citenamefont {Brault}, \citenamefont {Chenot}, \citenamefont
  {Dussaigne}, \citenamefont {Leroux},\ and\ \citenamefont
  {Damilano}}]{brochen2013dependence}%
  \BibitemOpen
  \bibfield  {author} {\bibinfo {author} {\bibfnamefont {S.}~\bibnamefont
  {Brochen}}, \bibinfo {author} {\bibfnamefont {J.}~\bibnamefont {Brault}},
  \bibinfo {author} {\bibfnamefont {S.}~\bibnamefont {Chenot}}, \bibinfo
  {author} {\bibfnamefont {A.}~\bibnamefont {Dussaigne}}, \bibinfo {author}
  {\bibfnamefont {M.}~\bibnamefont {Leroux}}, \ and\ \bibinfo {author}
  {\bibfnamefont {B.}~\bibnamefont {Damilano}},\ }\href@noop {} {\bibfield
  {journal} {\bibinfo  {journal} {Appl. Phys. Lett.}\ }\textbf {\bibinfo
  {volume} {103}},\ \bibinfo {pages} {032102} (\bibinfo {year}
  {2013})}\BibitemShut {NoStop}%
\bibitem [{\citenamefont {Castiglia}\ \emph {et~al.}(2011)\citenamefont
  {Castiglia}, \citenamefont {Carlin},\ and\ \citenamefont
  {Grandjean}}]{castiglia2011role}%
  \BibitemOpen
  \bibfield  {author} {\bibinfo {author} {\bibfnamefont {A.}~\bibnamefont
  {Castiglia}}, \bibinfo {author} {\bibfnamefont {J.-F.}\ \bibnamefont
  {Carlin}}, \ and\ \bibinfo {author} {\bibfnamefont {N.}~\bibnamefont
  {Grandjean}},\ }\href@noop {} {\bibfield  {journal} {\bibinfo  {journal}
  {Appl. Phys. Lett.}\ }\textbf {\bibinfo {volume} {98}},\ \bibinfo {pages}
  {213505} (\bibinfo {year} {2011})}\BibitemShut {NoStop}%
\bibitem [{\citenamefont {Liliental-Weber}\ \emph {et~al.}(1999)\citenamefont
  {Liliental-Weber}, \citenamefont {Benamara}, \citenamefont {Swider},
  \citenamefont {Washburn}, \citenamefont {Grzegory}, \citenamefont {Porowski},
  \citenamefont {Lambert}, \citenamefont {Eiting},\ and\ \citenamefont
  {Dupuis}}]{PID_APL99}%
  \BibitemOpen
  \bibfield  {author} {\bibinfo {author} {\bibfnamefont {Z.}~\bibnamefont
  {Liliental-Weber}}, \bibinfo {author} {\bibfnamefont {M.}~\bibnamefont
  {Benamara}}, \bibinfo {author} {\bibfnamefont {W.}~\bibnamefont {Swider}},
  \bibinfo {author} {\bibfnamefont {J.}~\bibnamefont {Washburn}}, \bibinfo
  {author} {\bibfnamefont {I.}~\bibnamefont {Grzegory}}, \bibinfo {author}
  {\bibfnamefont {S.}~\bibnamefont {Porowski}}, \bibinfo {author}
  {\bibfnamefont {D.~J.~H.}\ \bibnamefont {Lambert}}, \bibinfo {author}
  {\bibfnamefont {C.~J.}\ \bibnamefont {Eiting}}, \ and\ \bibinfo {author}
  {\bibfnamefont {R.~D.}\ \bibnamefont {Dupuis}},\ }\href {\doibase
  10.1063/1.125568} {\bibfield  {journal} {\bibinfo  {journal} {Appl. Phys.
  Lett.}\ }\textbf {\bibinfo {volume} {75}},\ \bibinfo {pages} {4159} (\bibinfo
  {year} {1999})}\BibitemShut {NoStop}%
\bibitem [{\citenamefont {Yan}\ \emph {et~al.}(2012)\citenamefont {Yan},
  \citenamefont {Janotti}, \citenamefont {Scheffler},\ and\ \citenamefont
  {Van~de Walle}}]{WalleAPL12}%
  \BibitemOpen
  \bibfield  {author} {\bibinfo {author} {\bibfnamefont {Q.}~\bibnamefont
  {Yan}}, \bibinfo {author} {\bibfnamefont {A.}~\bibnamefont {Janotti}},
  \bibinfo {author} {\bibfnamefont {M.}~\bibnamefont {Scheffler}}, \ and\
  \bibinfo {author} {\bibfnamefont {C.~G.}\ \bibnamefont {Van~de Walle}},\
  }\href {\doibase 10.1063/1.3699009} {\bibfield  {journal} {\bibinfo
  {journal} {Appl. Phys. Lett.}\ }\textbf {\bibinfo {volume} {100}},\ \bibinfo
  {pages} {142110} (\bibinfo {year} {2012})}\BibitemShut {NoStop}%
\bibitem [{\citenamefont {Miceli}\ and\ \citenamefont
  {Pasquarello}(2016)}]{PhysRevB.93.165207}%
  \BibitemOpen
  \bibfield  {author} {\bibinfo {author} {\bibfnamefont {G.}~\bibnamefont
  {Miceli}}\ and\ \bibinfo {author} {\bibfnamefont {A.}~\bibnamefont
  {Pasquarello}},\ }\href {\doibase 10.1103/PhysRevB.93.165207} {\bibfield
  {journal} {\bibinfo  {journal} {Phys. Rev. B}\ }\textbf {\bibinfo {volume}
  {93}},\ \bibinfo {pages} {165207} (\bibinfo {year} {2016})}\BibitemShut
  {NoStop}%
\bibitem [{\citenamefont {Lyons}\ \emph {et~al.}(2014)\citenamefont {Lyons},
  \citenamefont {Janotti},\ and\ \citenamefont {Van~de
  Walle}}]{lyons2014effects}%
  \BibitemOpen
  \bibfield  {author} {\bibinfo {author} {\bibfnamefont {J.}~\bibnamefont
  {Lyons}}, \bibinfo {author} {\bibfnamefont {A.}~\bibnamefont {Janotti}}, \
  and\ \bibinfo {author} {\bibfnamefont {C.}~\bibnamefont {Van~de Walle}},\
  }\href@noop {} {\bibfield  {journal} {\bibinfo  {journal} {Phys. Rev. B}\
  }\textbf {\bibinfo {volume} {89}},\ \bibinfo {pages} {035204} (\bibinfo
  {year} {2014})}\BibitemShut {NoStop}%
\bibitem [{\citenamefont {Klump}\ \emph {et~al.}(2020)\citenamefont {Klump},
  \citenamefont {Hoffmann}, \citenamefont {Kaess}, \citenamefont {Tweedie},
  \citenamefont {Reddy}, \citenamefont {Kirste}, \citenamefont {Sitar},\ and\
  \citenamefont {Collazo}}]{klump2020control}%
  \BibitemOpen
  \bibfield  {author} {\bibinfo {author} {\bibfnamefont {A.}~\bibnamefont
  {Klump}}, \bibinfo {author} {\bibfnamefont {M.}~\bibnamefont {Hoffmann}},
  \bibinfo {author} {\bibfnamefont {F.}~\bibnamefont {Kaess}}, \bibinfo
  {author} {\bibfnamefont {J.}~\bibnamefont {Tweedie}}, \bibinfo {author}
  {\bibfnamefont {P.}~\bibnamefont {Reddy}}, \bibinfo {author} {\bibfnamefont
  {R.}~\bibnamefont {Kirste}}, \bibinfo {author} {\bibfnamefont
  {Z.}~\bibnamefont {Sitar}}, \ and\ \bibinfo {author} {\bibfnamefont
  {R.}~\bibnamefont {Collazo}},\ }\href@noop {} {\bibfield  {journal} {\bibinfo
   {journal} {J. Appl. Phys.}\ }\textbf {\bibinfo {volume} {127}},\ \bibinfo
  {pages} {045702} (\bibinfo {year} {2020})}\BibitemShut {NoStop}%
\bibitem [{\citenamefont {Kakanakova-Georgieva}\ \emph
  {et~al.}(2009)\citenamefont {Kakanakova-Georgieva}, \citenamefont
  {Ciechonski}, \citenamefont {Forsberg}, \citenamefont {Lundskog},\ and\
  \citenamefont {Janz{\'{e}}n}}]{Kakanakova-Georgieva2009Hot-wallAlN}%
  \BibitemOpen
  \bibfield  {author} {\bibinfo {author} {\bibfnamefont {A.}~\bibnamefont
  {Kakanakova-Georgieva}}, \bibinfo {author} {\bibfnamefont {R.~R.}\
  \bibnamefont {Ciechonski}}, \bibinfo {author} {\bibfnamefont
  {U.}~\bibnamefont {Forsberg}}, \bibinfo {author} {\bibfnamefont
  {A.}~\bibnamefont {Lundskog}}, \ and\ \bibinfo {author} {\bibfnamefont
  {E.}~\bibnamefont {Janz{\'{e}}n}},\ }\href {\doibase 10.1021/cg8005663}
  {\bibfield  {journal} {\bibinfo  {journal} {Cryst. Growth Des.}\ }\textbf
  {\bibinfo {volume} {9}},\ \bibinfo {pages} {880} (\bibinfo {year}
  {2009})}\BibitemShut {NoStop}%
\bibitem [{\citenamefont {Sch{\"o}che}\ \emph {et~al.}(2017)\citenamefont
  {Sch{\"o}che}, \citenamefont {Hofmann}, \citenamefont {Nilsson},
  \citenamefont {Kakanakova-Georgieva}, \citenamefont {Janz{\'{e}}n},
  \citenamefont {K{\"{u}}hne}, \citenamefont {Lorenz}, \citenamefont
  {Schubert},\ and\ \citenamefont {Darakchieva}}]{Schoche2017InfraredEffect}%
  \BibitemOpen
  \bibfield  {author} {\bibinfo {author} {\bibfnamefont {S.}~\bibnamefont
  {Sch{\"o}che}}, \bibinfo {author} {\bibfnamefont {T.}~\bibnamefont
  {Hofmann}}, \bibinfo {author} {\bibfnamefont {D.}~\bibnamefont {Nilsson}},
  \bibinfo {author} {\bibfnamefont {A.}~\bibnamefont {Kakanakova-Georgieva}},
  \bibinfo {author} {\bibfnamefont {E.}~\bibnamefont {Janz{\'{e}}n}}, \bibinfo
  {author} {\bibfnamefont {P.}~\bibnamefont {K{\"{u}}hne}}, \bibinfo {author}
  {\bibfnamefont {K.}~\bibnamefont {Lorenz}}, \bibinfo {author} {\bibfnamefont
  {M.}~\bibnamefont {Schubert}}, \ and\ \bibinfo {author} {\bibfnamefont
  {V.}~\bibnamefont {Darakchieva}},\ }\href {\doibase 10.1063/1.4983765}
  {\bibfield  {journal} {\bibinfo  {journal} {J. Appl. Phys.}\ }\textbf
  {\bibinfo {volume} {121}},\ \bibinfo {pages} {205701} (\bibinfo {year}
  {2017})}\BibitemShut {NoStop}%
\bibitem [{\citenamefont {Feneberg}\ \emph {et~al.}(2015)\citenamefont
  {Feneberg}, \citenamefont {Son},\ and\ \citenamefont
  {Kakanakova-Georgieva}}]{Feneberg2015ExcitonAnnealing}%
  \BibitemOpen
  \bibfield  {author} {\bibinfo {author} {\bibfnamefont {M.}~\bibnamefont
  {Feneberg}}, \bibinfo {author} {\bibfnamefont {N.~T.}\ \bibnamefont {Son}}, \
  and\ \bibinfo {author} {\bibfnamefont {A.}~\bibnamefont
  {Kakanakova-Georgieva}},\ }\href {\doibase 10.1063/1.4922723} {\bibfield
  {journal} {\bibinfo  {journal} {Appl. Phys. Lett.}\ }\textbf {\bibinfo
  {volume} {106}},\ \bibinfo {pages} {242101} (\bibinfo {year}
  {2015})}\BibitemShut {NoStop}%
\bibitem [{\citenamefont {Kakanakova-Georgieva}\ \emph
  {et~al.}(2010)\citenamefont {Kakanakova-Georgieva}, \citenamefont {Nilsson},
  \citenamefont {Stattin}, \citenamefont {Forsberg}, \citenamefont {Haglund},
  \citenamefont {Larsson},\ and\ \citenamefont
  {Janz{\'e}n}}]{Kakanakova-Georgieva2010Mg-dopedTemperature}%
  \BibitemOpen
  \bibfield  {author} {\bibinfo {author} {\bibfnamefont {A.}~\bibnamefont
  {Kakanakova-Georgieva}}, \bibinfo {author} {\bibfnamefont {D.}~\bibnamefont
  {Nilsson}}, \bibinfo {author} {\bibfnamefont {M.}~\bibnamefont {Stattin}},
  \bibinfo {author} {\bibfnamefont {U.}~\bibnamefont {Forsberg}}, \bibinfo
  {author} {\bibfnamefont {A.}~\bibnamefont {Haglund}}, \bibinfo {author}
  {\bibfnamefont {A.}~\bibnamefont {Larsson}}, \ and\ \bibinfo {author}
  {\bibfnamefont {E.}~\bibnamefont {Janz{\'e}n}},\ }\href {\doibase
  10.1002/pssr.201004290} {\bibfield  {journal} {\bibinfo  {journal} {Phys.
  Status Solidi Rapid Res. Lett.}\ }\textbf {\bibinfo {volume} {4}},\ \bibinfo
  {pages} {311} (\bibinfo {year} {2010})}\BibitemShut {NoStop}%
\bibitem [{\citenamefont {Chen}\ \emph {et~al.}(2015)\citenamefont {Chen},
  \citenamefont {Persson}, \citenamefont {Nilsson}, \citenamefont {Hsu},
  \citenamefont {Palisaitis}, \citenamefont {Forsberg}, \citenamefont
  {Persson},\ and\ \citenamefont
  {Janz{\'{e}}n}}]{Chen2015Room-temperatureHeterostructure}%
  \BibitemOpen
  \bibfield  {author} {\bibinfo {author} {\bibfnamefont {J.~T.}\ \bibnamefont
  {Chen}}, \bibinfo {author} {\bibfnamefont {I.}~\bibnamefont {Persson}},
  \bibinfo {author} {\bibfnamefont {D.}~\bibnamefont {Nilsson}}, \bibinfo
  {author} {\bibfnamefont {C.~W.}\ \bibnamefont {Hsu}}, \bibinfo {author}
  {\bibfnamefont {J.}~\bibnamefont {Palisaitis}}, \bibinfo {author}
  {\bibfnamefont {U.}~\bibnamefont {Forsberg}}, \bibinfo {author}
  {\bibfnamefont {P.~O.}\ \bibnamefont {Persson}}, \ and\ \bibinfo {author}
  {\bibfnamefont {E.}~\bibnamefont {Janz{\'{e}}n}},\ }\href {\doibase
  10.1063/1.4922877} {\bibfield  {journal} {\bibinfo  {journal} {Appl. Phys.
  Lett.}\ }\textbf {\bibinfo {volume} {106}},\ \bibinfo {pages} {251601}
  (\bibinfo {year} {2015})}\BibitemShut {NoStop}%
\bibitem [{\citenamefont {Armakavicius}\ \emph {et~al.}(2016)\citenamefont
  {Armakavicius}, \citenamefont {Chen}, \citenamefont {Hofmann}, \citenamefont
  {Knight}, \citenamefont {K{\"u}hne}, \citenamefont {Nilsson}, \citenamefont
  {Forsberg}, \citenamefont {Janz{\'{e}}n},\ and\ \citenamefont
  {Darakchieva}}]{Armakavicius2016PropertiesEffect}%
  \BibitemOpen
  \bibfield  {author} {\bibinfo {author} {\bibfnamefont {N.}~\bibnamefont
  {Armakavicius}}, \bibinfo {author} {\bibfnamefont {J.-T.}\ \bibnamefont
  {Chen}}, \bibinfo {author} {\bibfnamefont {T.}~\bibnamefont {Hofmann}},
  \bibinfo {author} {\bibfnamefont {S.}~\bibnamefont {Knight}}, \bibinfo
  {author} {\bibfnamefont {P.}~\bibnamefont {K{\"u}hne}}, \bibinfo {author}
  {\bibfnamefont {D.}~\bibnamefont {Nilsson}}, \bibinfo {author} {\bibfnamefont
  {U.}~\bibnamefont {Forsberg}}, \bibinfo {author} {\bibfnamefont
  {E.}~\bibnamefont {Janz{\'{e}}n}}, \ and\ \bibinfo {author} {\bibfnamefont
  {V.}~\bibnamefont {Darakchieva}},\ }\href {\doibase 10.1002/pssc.201510214}
  {\bibfield  {journal} {\bibinfo  {journal} {Phys. Status Solidi C}\ }\textbf
  {\bibinfo {volume} {13}},\ \bibinfo {pages} {369} (\bibinfo {year}
  {2016})}\BibitemShut {NoStop}%
\bibitem [{\citenamefont {Chen}\ \emph {et~al.}(2018)\citenamefont {Chen},
  \citenamefont {Bergsten}, \citenamefont {Lu}, \citenamefont {Janz{\'{e}}n},
  \citenamefont {Thorsell}, \citenamefont {Hultman}, \citenamefont {Rorsman},\
  and\ \citenamefont {Kordina}}]{Chen2018AElectronics}%
  \BibitemOpen
  \bibfield  {author} {\bibinfo {author} {\bibfnamefont {J.~T.}\ \bibnamefont
  {Chen}}, \bibinfo {author} {\bibfnamefont {J.}~\bibnamefont {Bergsten}},
  \bibinfo {author} {\bibfnamefont {J.}~\bibnamefont {Lu}}, \bibinfo {author}
  {\bibfnamefont {E.}~\bibnamefont {Janz{\'{e}}n}}, \bibinfo {author}
  {\bibfnamefont {M.}~\bibnamefont {Thorsell}}, \bibinfo {author}
  {\bibfnamefont {L.}~\bibnamefont {Hultman}}, \bibinfo {author} {\bibfnamefont
  {N.}~\bibnamefont {Rorsman}}, \ and\ \bibinfo {author} {\bibfnamefont
  {O.}~\bibnamefont {Kordina}},\ }\href {\doibase 10.1063/1.5042049} {\bibfield
   {journal} {\bibinfo  {journal} {Appl. Phys. Lett.}\ }\textbf {\bibinfo
  {volume} {113}},\ \bibinfo {pages} {041605} (\bibinfo {year}
  {2018})}\BibitemShut {NoStop}%
\bibitem [{\citenamefont {Zhang}\ \emph {et~al.}(2020)\citenamefont {Zhang},
  \citenamefont {Paskov}, \citenamefont {Kordina}, \citenamefont {Chen},\ and\
  \citenamefont {Darakchieva}}]{Zhang_2020n}%
  \BibitemOpen
  \bibfield  {author} {\bibinfo {author} {\bibfnamefont {H.}~\bibnamefont
  {Zhang}}, \bibinfo {author} {\bibfnamefont {P.~P.}\ \bibnamefont {Paskov}},
  \bibinfo {author} {\bibfnamefont {O.}~\bibnamefont {Kordina}}, \bibinfo
  {author} {\bibfnamefont {J.-T.}\ \bibnamefont {Chen}}, \ and\ \bibinfo
  {author} {\bibfnamefont {V.}~\bibnamefont {Darakchieva}},\ }\href@noop {}
  {\bibfield  {journal} {\bibinfo  {journal} {Physica B: Condens. Matter}\
  }\textbf {\bibinfo {volume} {580}},\ \bibinfo {pages} {411819} (\bibinfo
  {year} {2020})}\BibitemShut {NoStop}%
\bibitem [{EAG()}]{EAGLaboratoriesHttps://www.eag.com}%
  \BibitemOpen
  \href@noop {} {}\bibinfo {note} {{EAG Laboratories
  (https://www.eag.com)}}\BibitemShut {NoStop}%
\bibitem [{\citenamefont {Srikant}\ \emph {et~al.}(1997)\citenamefont
  {Srikant}, \citenamefont {Speck},\ and\ \citenamefont
  {Clarke}}]{Srikant1997MosaicMismatch}%
  \BibitemOpen
  \bibfield  {author} {\bibinfo {author} {\bibfnamefont {V.}~\bibnamefont
  {Srikant}}, \bibinfo {author} {\bibfnamefont {J.~S.}\ \bibnamefont {Speck}},
  \ and\ \bibinfo {author} {\bibfnamefont {D.~R.}\ \bibnamefont {Clarke}},\
  }\href@noop {} {\bibfield  {journal} {\bibinfo  {journal} {J. Appl. Phys.}\
  }\textbf {\bibinfo {volume} {82}},\ \bibinfo {pages} {4286} (\bibinfo {year}
  {1997})}\BibitemShut {NoStop}%
\bibitem [{\citenamefont {Metzger}\ \emph {et~al.}(1998)\citenamefont
  {Metzger}, \citenamefont {H{\"{o}}pler}, \citenamefont {Born}, \citenamefont
  {Ambacher}, \citenamefont {Stutzmann}, \citenamefont {St{\"{o}}mmer},
  \citenamefont {Schuster}, \citenamefont {G{\"{o}}bel}, \citenamefont
  {Christiansen}, \citenamefont {Albrecht},\ and\ \citenamefont
  {Strunk}}]{Metzger1998DefectDiffractometry}%
  \BibitemOpen
  \bibfield  {author} {\bibinfo {author} {\bibfnamefont {T.}~\bibnamefont
  {Metzger}}, \bibinfo {author} {\bibfnamefont {R.}~\bibnamefont
  {H{\"{o}}pler}}, \bibinfo {author} {\bibfnamefont {E.}~\bibnamefont {Born}},
  \bibinfo {author} {\bibfnamefont {O.}~\bibnamefont {Ambacher}}, \bibinfo
  {author} {\bibfnamefont {M.}~\bibnamefont {Stutzmann}}, \bibinfo {author}
  {\bibfnamefont {R.}~\bibnamefont {St{\"{o}}mmer}}, \bibinfo {author}
  {\bibfnamefont {M.}~\bibnamefont {Schuster}}, \bibinfo {author}
  {\bibfnamefont {H.}~\bibnamefont {G{\"{o}}bel}}, \bibinfo {author}
  {\bibfnamefont {S.}~\bibnamefont {Christiansen}}, \bibinfo {author}
  {\bibfnamefont {M.}~\bibnamefont {Albrecht}}, \ and\ \bibinfo {author}
  {\bibfnamefont {H.~P.}\ \bibnamefont {Strunk}},\ }\href@noop {} {\bibfield
  {journal} {\bibinfo  {journal} {Philos. Mag. A}\ }\textbf {\bibinfo {volume}
  {77}},\ \bibinfo {pages} {1013} (\bibinfo {year} {1998})}\BibitemShut
  {NoStop}%
\bibitem [{\citenamefont {Darakchieva}\ \emph {et~al.}(2007)\citenamefont
  {Darakchieva}, \citenamefont {Monemar},\ and\ \citenamefont
  {Usui}}]{Darakchieva_2007}%
  \BibitemOpen
  \bibfield  {author} {\bibinfo {author} {\bibfnamefont {V.}~\bibnamefont
  {Darakchieva}}, \bibinfo {author} {\bibfnamefont {B.}~\bibnamefont
  {Monemar}}, \ and\ \bibinfo {author} {\bibfnamefont {A.}~\bibnamefont
  {Usui}},\ }\href@noop {} {\bibfield  {journal} {\bibinfo  {journal} {Appl.
  Phys. Lett.}\ }\textbf {\bibinfo {volume} {91}},\ \bibinfo {pages} {031911}
  (\bibinfo {year} {2007})}\BibitemShut {NoStop}%
\bibitem [{\citenamefont {Tokunaga}\ \emph {et~al.}(1998)\citenamefont
  {Tokunaga}, \citenamefont {Waki}, \citenamefont {Yamaguchi}, \citenamefont
  {Akutsu},\ and\ \citenamefont {Matsumoto}}]{tokunaga1998growth}%
  \BibitemOpen
  \bibfield  {author} {\bibinfo {author} {\bibfnamefont {H.}~\bibnamefont
  {Tokunaga}}, \bibinfo {author} {\bibfnamefont {I.}~\bibnamefont {Waki}},
  \bibinfo {author} {\bibfnamefont {A.}~\bibnamefont {Yamaguchi}}, \bibinfo
  {author} {\bibfnamefont {N.}~\bibnamefont {Akutsu}}, \ and\ \bibinfo {author}
  {\bibfnamefont {K.}~\bibnamefont {Matsumoto}},\ }\href@noop {} {\bibfield
  {journal} {\bibinfo  {journal} {J. Cryst. Growth}\ }\textbf {\bibinfo
  {volume} {189}},\ \bibinfo {pages} {519} (\bibinfo {year}
  {1998})}\BibitemShut {NoStop}%
\bibitem [{\citenamefont {De~Mierry}\ \emph {et~al.}(2000)\citenamefont
  {De~Mierry}, \citenamefont {Beaumont}, \citenamefont {Feltin}, \citenamefont
  {Schenk}, \citenamefont {Gibart}, \citenamefont {Jomard}, \citenamefont
  {Rushworth}, \citenamefont {Smith},\ and\ \citenamefont
  {Odedra}}]{de2000influence}%
  \BibitemOpen
  \bibfield  {author} {\bibinfo {author} {\bibfnamefont {P.}~\bibnamefont
  {De~Mierry}}, \bibinfo {author} {\bibfnamefont {B.}~\bibnamefont {Beaumont}},
  \bibinfo {author} {\bibfnamefont {E.}~\bibnamefont {Feltin}}, \bibinfo
  {author} {\bibfnamefont {H.}~\bibnamefont {Schenk}}, \bibinfo {author}
  {\bibfnamefont {P.}~\bibnamefont {Gibart}}, \bibinfo {author} {\bibfnamefont
  {F.}~\bibnamefont {Jomard}}, \bibinfo {author} {\bibfnamefont
  {S.}~\bibnamefont {Rushworth}}, \bibinfo {author} {\bibfnamefont
  {L.}~\bibnamefont {Smith}}, \ and\ \bibinfo {author} {\bibfnamefont
  {R.}~\bibnamefont {Odedra}},\ }\href@noop {} {\bibfield  {journal} {\bibinfo
  {journal} {MRS Internet J. Nitride Semicond. Res.}\ }\textbf {\bibinfo
  {volume} {5}},\ \bibinfo {pages} {1} (\bibinfo {year} {2000})}\BibitemShut
  {NoStop}%
\bibitem [{\citenamefont {Kakanakova-Georgieva}\ \emph {et~al.}()\citenamefont
  {Kakanakova-Georgieva}, \citenamefont {Papamichail}, \citenamefont
  {Stanishev},\ and\ \citenamefont {Darakchieva}}]{AneliaAPL22}%
  \BibitemOpen
  \bibfield  {author} {\bibinfo {author} {\bibfnamefont {A.}~\bibnamefont
  {Kakanakova-Georgieva}}, \bibinfo {author} {\bibfnamefont {A.}~\bibnamefont
  {Papamichail}}, \bibinfo {author} {\bibfnamefont {V.}~\bibnamefont
  {Stanishev}}, \ and\ \bibinfo {author} {\bibfnamefont {V.}~\bibnamefont
  {Darakchieva}},\ }\href@noop {} {\ }\bibinfo {note}
  {(unpublished)}\BibitemShut {NoStop}%
\bibitem [{\citenamefont {Mita}\ \emph {et~al.}(2008)\citenamefont {Mita},
  \citenamefont {Collazo}, \citenamefont {Rice}, \citenamefont {Dalmau},\ and\
  \citenamefont {Sitar}}]{MitaJAP08}%
  \BibitemOpen
  \bibfield  {author} {\bibinfo {author} {\bibfnamefont {S.}~\bibnamefont
  {Mita}}, \bibinfo {author} {\bibfnamefont {R.}~\bibnamefont {Collazo}},
  \bibinfo {author} {\bibfnamefont {A.}~\bibnamefont {Rice}}, \bibinfo {author}
  {\bibfnamefont {R.~F.}\ \bibnamefont {Dalmau}}, \ and\ \bibinfo {author}
  {\bibfnamefont {Z.}~\bibnamefont {Sitar}},\ }\href {\doibase
  10.1063/1.2952027} {\bibfield  {journal} {\bibinfo  {journal} {J. Appl.
  Phys.}\ }\textbf {\bibinfo {volume} {104}},\ \bibinfo {pages} {013521}
  (\bibinfo {year} {2008})}\BibitemShut {NoStop}%
\bibitem [{Note1()}]{Note1}%
  \BibitemOpen
  \bibinfo {note} {Samples M$_2$ and M$_5$ were grown with fresh satellite,
  which could explain the slight variations in comparison to samples M$_1$,
  M$_3$ and M$_4$. We note that this has no baring on the reported general
  trends and conclusions.}\BibitemShut {Stop}%
\bibitem [{\citenamefont {Xie}\ \emph {et~al.}(2014)\citenamefont {Xie},
  \citenamefont {Sedrine.}, \citenamefont {Schoche}, \citenamefont {Hofmann},
  \citenamefont {Schubert}, \citenamefont {Hung}, \citenamefont {Monemar},
  \citenamefont {Wang}, \citenamefont {Yoshikawa}, \citenamefont {Wang},
  \citenamefont {Araki}, \citenamefont {Nanishi},\ and\ \citenamefont
  {Darakchieva}}]{XieJAP14}%
  \BibitemOpen
  \bibfield  {author} {\bibinfo {author} {\bibfnamefont {M.-Y.}\ \bibnamefont
  {Xie}}, \bibinfo {author} {\bibfnamefont {N.~B.}\ \bibnamefont {Sedrine.}},
  \bibinfo {author} {\bibfnamefont {S.}~\bibnamefont {Schoche}}, \bibinfo
  {author} {\bibfnamefont {T.}~\bibnamefont {Hofmann}}, \bibinfo {author}
  {\bibfnamefont {M.}~\bibnamefont {Schubert}}, \bibinfo {author}
  {\bibfnamefont {L.}~\bibnamefont {Hung}}, \bibinfo {author} {\bibfnamefont
  {B.}~\bibnamefont {Monemar}}, \bibinfo {author} {\bibfnamefont
  {X.}~\bibnamefont {Wang}}, \bibinfo {author} {\bibfnamefont {A.}~\bibnamefont
  {Yoshikawa}}, \bibinfo {author} {\bibfnamefont {K.}~\bibnamefont {Wang}},
  \bibinfo {author} {\bibfnamefont {T.}~\bibnamefont {Araki}}, \bibinfo
  {author} {\bibfnamefont {Y.}~\bibnamefont {Nanishi}}, \ and\ \bibinfo
  {author} {\bibfnamefont {V.}~\bibnamefont {Darakchieva}},\ }\href@noop {}
  {\bibfield  {journal} {\bibinfo  {journal} {J. Appl. Phys.}\ }\textbf
  {\bibinfo {volume} {115}},\ \bibinfo {pages} {163504} (\bibinfo {year}
  {2014})}\BibitemShut {NoStop}%
\bibitem [{\citenamefont {Ramachandran}\ \emph {et~al.}(1999)\citenamefont
  {Ramachandran}, \citenamefont {Feenstra}, \citenamefont {Sarney},
  \citenamefont {Salamanca-Riba}, \citenamefont {Northrup}, \citenamefont
  {Romano},\ and\ \citenamefont {Greve}}]{RomanoAPL99}%
  \BibitemOpen
  \bibfield  {author} {\bibinfo {author} {\bibfnamefont {V.}~\bibnamefont
  {Ramachandran}}, \bibinfo {author} {\bibfnamefont {R.~M.}\ \bibnamefont
  {Feenstra}}, \bibinfo {author} {\bibfnamefont {W.~L.}\ \bibnamefont
  {Sarney}}, \bibinfo {author} {\bibfnamefont {L.}~\bibnamefont
  {Salamanca-Riba}}, \bibinfo {author} {\bibfnamefont {J.~E.}\ \bibnamefont
  {Northrup}}, \bibinfo {author} {\bibfnamefont {L.~T.}\ \bibnamefont
  {Romano}}, \ and\ \bibinfo {author} {\bibfnamefont {D.~W.}\ \bibnamefont
  {Greve}},\ }\href {\doibase 10.1063/1.124520} {\bibfield  {journal} {\bibinfo
   {journal} {Appl. Phys. Lett.}\ }\textbf {\bibinfo {volume} {75}},\ \bibinfo
  {pages} {808} (\bibinfo {year} {1999})}\BibitemShut {NoStop}%
\bibitem [{\citenamefont {Tavernier}\ \emph {et~al.}(2004)\citenamefont
  {Tavernier}, \citenamefont {Margalith}, \citenamefont {Williams},
  \citenamefont {Green}, \citenamefont {Keller}, \citenamefont {DenBaars},
  \citenamefont {Mishra}, \citenamefont {Nakamura},\ and\ \citenamefont
  {Clarke}}]{Tavernier2004TheIn}%
  \BibitemOpen
  \bibfield  {author} {\bibinfo {author} {\bibfnamefont {P.~R.}\ \bibnamefont
  {Tavernier}}, \bibinfo {author} {\bibfnamefont {T.}~\bibnamefont
  {Margalith}}, \bibinfo {author} {\bibfnamefont {J.}~\bibnamefont {Williams}},
  \bibinfo {author} {\bibfnamefont {D.~S.}\ \bibnamefont {Green}}, \bibinfo
  {author} {\bibfnamefont {S.}~\bibnamefont {Keller}}, \bibinfo {author}
  {\bibfnamefont {S.~P.}\ \bibnamefont {DenBaars}}, \bibinfo {author}
  {\bibfnamefont {U.~K.}\ \bibnamefont {Mishra}}, \bibinfo {author}
  {\bibfnamefont {S.}~\bibnamefont {Nakamura}}, \ and\ \bibinfo {author}
  {\bibfnamefont {D.~R.}\ \bibnamefont {Clarke}},\ }\href {\doibase
  10.1016/j.jcrysgro.2004.01.023} {\bibfield  {journal} {\bibinfo  {journal}
  {J. Cryst. Growth}\ }\textbf {\bibinfo {volume} {264}},\ \bibinfo {pages}
  {150} (\bibinfo {year} {2004})}\BibitemShut {NoStop}%
\bibitem [{\citenamefont {Fichtenbaum}\ \emph {et~al.}(2007)\citenamefont
  {Fichtenbaum}, \citenamefont {Schaake}, \citenamefont {Mates}, \citenamefont
  {Cobb}, \citenamefont {Keller}, \citenamefont {DenBaars},\ and\ \citenamefont
  {Mishra}}]{N-polarMgdopedGaN-APL07}%
  \BibitemOpen
  \bibfield  {author} {\bibinfo {author} {\bibfnamefont {N.~A.}\ \bibnamefont
  {Fichtenbaum}}, \bibinfo {author} {\bibfnamefont {C.}~\bibnamefont
  {Schaake}}, \bibinfo {author} {\bibfnamefont {T.~E.}\ \bibnamefont {Mates}},
  \bibinfo {author} {\bibfnamefont {C.}~\bibnamefont {Cobb}}, \bibinfo {author}
  {\bibfnamefont {S.}~\bibnamefont {Keller}}, \bibinfo {author} {\bibfnamefont
  {S.~P.}\ \bibnamefont {DenBaars}}, \ and\ \bibinfo {author} {\bibfnamefont
  {U.~K.}\ \bibnamefont {Mishra}},\ }\href {\doibase 10.1063/1.2800304}
  {\bibfield  {journal} {\bibinfo  {journal} {Appl. Phys. Lett.}\ }\textbf
  {\bibinfo {volume} {91}},\ \bibinfo {pages} {172105} (\bibinfo {year}
  {2007})}\BibitemShut {NoStop}%
\bibitem [{\citenamefont {Neugebauer}\ and\ \citenamefont {Van~de
  Walle}(1996)}]{WalleAPL96}%
  \BibitemOpen
  \bibfield  {author} {\bibinfo {author} {\bibfnamefont {J.}~\bibnamefont
  {Neugebauer}}\ and\ \bibinfo {author} {\bibfnamefont {C.~G.}\ \bibnamefont
  {Van~de Walle}},\ }\href@noop {} {\bibfield  {journal} {\bibinfo  {journal}
  {Appl. Phys. Lett.}\ }\textbf {\bibinfo {volume} {68}},\ \bibinfo {pages}
  {1829} (\bibinfo {year} {1996})}\BibitemShut {NoStop}%
\bibitem [{\citenamefont {Lyons}\ \emph {et~al.}(2012)\citenamefont {Lyons},
  \citenamefont {Janotti},\ and\ \citenamefont {Van~de Walle}}]{WallePRL12}%
  \BibitemOpen
  \bibfield  {author} {\bibinfo {author} {\bibfnamefont {J.~L.}\ \bibnamefont
  {Lyons}}, \bibinfo {author} {\bibfnamefont {A.}~\bibnamefont {Janotti}}, \
  and\ \bibinfo {author} {\bibfnamefont {C.~G.}\ \bibnamefont {Van~de Walle}},\
  }\href@noop {} {\bibfield  {journal} {\bibinfo  {journal} {Phys. Rev. Lett.}\
  }\textbf {\bibinfo {volume} {108}},\ \bibinfo {pages} {156403} (\bibinfo
  {year} {2012})}\BibitemShut {NoStop}%
\bibitem [{\citenamefont {Sang}\ \emph {et~al.}(2019)\citenamefont {Sang},
  \citenamefont {Ren}, \citenamefont {Endo}, \citenamefont {Masuda},
  \citenamefont {Yasufuku}, \citenamefont {Liao}, \citenamefont {Nabatame},
  \citenamefont {Sumiya},\ and\ \citenamefont {Koide}}]{Koide_APL19}%
  \BibitemOpen
  \bibfield  {author} {\bibinfo {author} {\bibfnamefont {L.}~\bibnamefont
  {Sang}}, \bibinfo {author} {\bibfnamefont {B.}~\bibnamefont {Ren}}, \bibinfo
  {author} {\bibfnamefont {R.}~\bibnamefont {Endo}}, \bibinfo {author}
  {\bibfnamefont {T.}~\bibnamefont {Masuda}}, \bibinfo {author} {\bibfnamefont
  {H.}~\bibnamefont {Yasufuku}}, \bibinfo {author} {\bibfnamefont
  {M.}~\bibnamefont {Liao}}, \bibinfo {author} {\bibfnamefont {T.}~\bibnamefont
  {Nabatame}}, \bibinfo {author} {\bibfnamefont {M.}~\bibnamefont {Sumiya}}, \
  and\ \bibinfo {author} {\bibfnamefont {Y.}~\bibnamefont {Koide}},\ }\href
  {\doibase 10.1063/1.5124904} {\bibfield  {journal} {\bibinfo  {journal}
  {Appl. Phys. Lett.}\ }\textbf {\bibinfo {volume} {115}},\ \bibinfo {pages}
  {172103} (\bibinfo {year} {2019})}\BibitemShut {NoStop}%
\end{thebibliography}%


\clearpage
\widetext

\section*{ \large{Supplementary: Mg-doping and free-hole properties of hot-wall MOCVD GaN}}

\fontsize{12}{24}\selectfont

\setcounter{section}{0}
\setcounter{equation}{0}
\setcounter{figure}{0}
\setcounter{table}{0}
\setcounter{page}{1}
\makeatletter

\renewcommand{\thefigure}{S\arabic{figure}}

 Figure S1 shows how the atomic fraction of Mg in GaN lattice varies as a function of the percentage of Cp$_2$/TMGa precursors flux in the gas phase. Mg concentration,[Mg], is measured by SIMS and [Ga]=$4.4\times{10}^{22}$ cm$^{-3}$. The slope K of the linear fitting shows that the Mg incorporation efficiency is K$\sim0.24$ and this value is very similar to literature reported values for p-type GaN growth in cold-wall MOCVD.

\begin{figure}[h!]
\includegraphics[keepaspectratio=true,width=0.6\textwidth, clip, trim=0cm 0cm 0cm 0cm ]{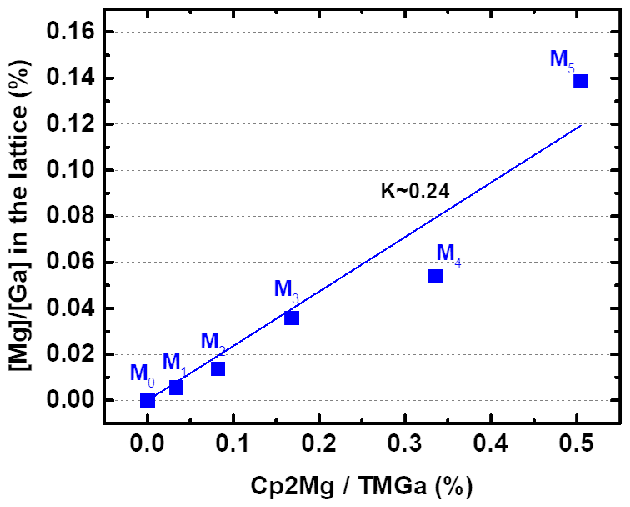}
\caption{Mg atomic fraction in GaN as a function of Cp$_2$/TMGa flux ratio in the gas phase. The slope K, of the linear fit, gives the Mg incorporation efficiency.}
\label{Mg incorp efficiency}
\end{figure}

\begin{figure}[h!]
\includegraphics[keepaspectratio=true,width=0.6\textwidth, clip, trim=0cm 0cm 0cm 0cm ]{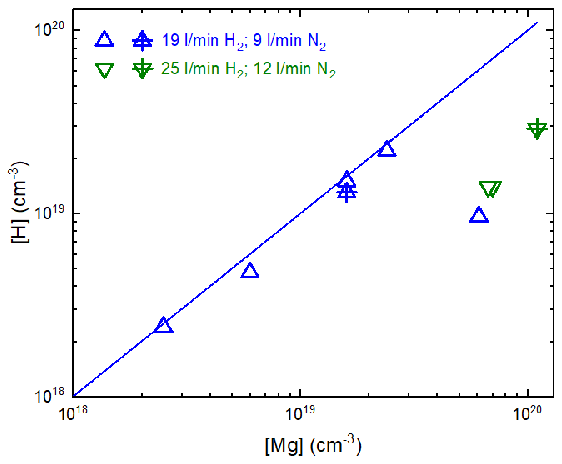}
\caption{Hydrogen concentration, [Mg] vs magnesium concentration, [Mg]. The solid like represents one to one correspondence between [Mg] and [H]. The same notation for the symbols representing samples grown at different growth conditions as in Fig. 1 is used.}
\label{HsvsMgMg}
\end{figure}

\begin{figure}[h!]
\includegraphics[keepaspectratio=true,width=0.6\textwidth, clip, trim=0cm 0cm 0cm 0cm ]{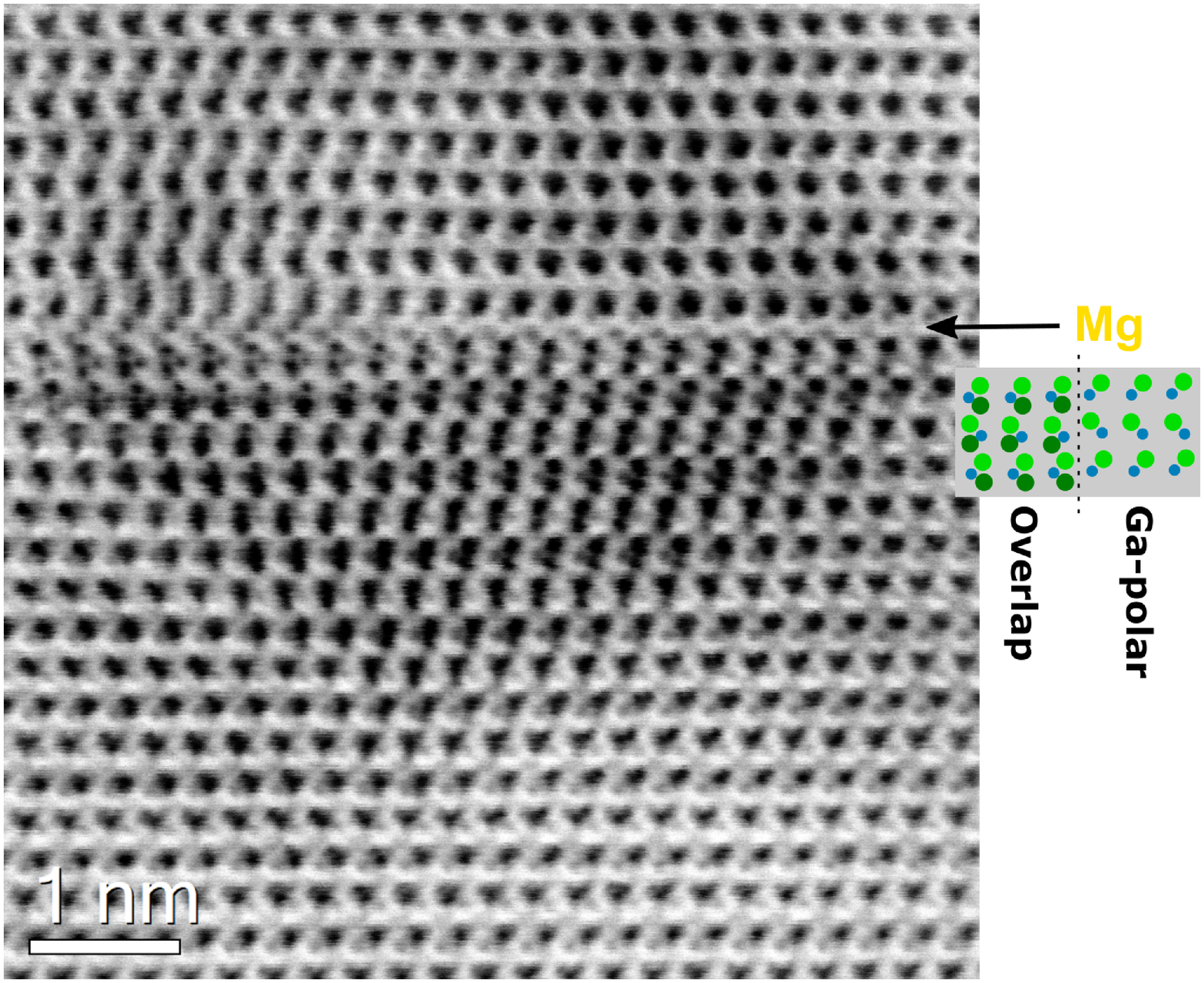}
\caption{Polarity inversion inside the pyramids. Image is acquired using annular bright-field (ABF) STEM using collection semi-angles of $\sim$4-43 mrad (resulting in an inverted contrast compared to HAADF, but also increased contrast of N). The layer of  Mg atoms segregating at the (0001) facet is marked with an arrow and the structure is shown schematically to the right. Growth direction [0001] is up. Ga is shown in green and N in blue (darker green for the N-polar structure in the pyramid). The N-polar pyramid and the Ga-polar matrix are seen overlapped.}
\label{polarity_TEM}
\end{figure}

\end{document}